\documentclass[12pt]{article}
\usepackage{graphics}
\usepackage[dvips]{graphicx}
\usepackage{mathrsfs}
\usepackage{amssymb}
\usepackage{verbatim}
\usepackage{float}
\newcommand{\be}{\begin{equation}}
\newcommand{\ee}{\end{equation}}
\newcommand{\bea}{\begin{eqnarray}}
\newcommand{\eea}{\end{eqnarray}}
\newcommand{\nnmb}{\nonumber}

\def\4vol{{\int d^4x \sqrt{-g}}}

\def\del{{\partial}}

\def\simgt{\stackrel{>}{{}_\sim}}
\def\su12{SU(2)_1\times SU(2)_2}

\def\sig{\Sigma_{i\bar{k}}}

\def\wt1{\tilde{W}_1^a}
\def\wt2{\tilde{W}_2^a}
\def\cf{c_{\varphi}}
\def\sf{s_{\varphi}}
\def\ct{c_{\theta}}
\def\st{s_{\theta}}
\newcommand{\nc}{\newcommand}

\nc{\avg}[1]{\left<#1\right>}
\nc{\bra}[1]{\left<#1\right|}
\nc{\braop}[1]{\left<#1\right.\!\!}
\nc{\ket}[1]{\left|#1\right>}

\nc{\ra}{\rightarrow}
\nc{\lsim}{\begin{array}{c}\,\sim\vspace{-21pt}\\< \end{array}}
\nc{\gsim}{\begin{array}{c}\sim\vspace{-21pt}\\> \end{array}}
\nc{\nt}{\tilde{N}}

\nc{\LL}{L}
\nc{\vv}{\mathcal{V}}
\nc{\ccdot}{\!\cdot\!}

\title{
\vspace*{-1.3cm}
\begin{flushright}
\normalsize{
ANL-HEP-PR-05-48\\
EFI-05-03
}
\end{flushright}
\vspace{1.5cm}
\Large
\textbf{Proton Lifetime and Baryon Number Violating
Signatures at the LHC in Gauge Extended Models}
\vspace*{1.0cm}
\author{\large 
\textbf{D.E.~Morrissey$^{a,b}$},
\textbf{T.M.P.~Tait$^{a}$}
and \textbf{C.E.M.~Wagner$^{a,b}$}\\ \\[0.5cm]
$^a$\normalsize\emph{HEP Division, Argonne National Laboratory,
9700 Cass Ave.,
Argonne, IL 60439, USA} \\
$^b$\normalsize\emph{Enrico Fermi Institute, University of Chicago, Chicago,
IL 60637, USA}}}

\begin{document}
\setcounter{page}{0}
\maketitle

%\date{}
\vspace*{1cm}
\begin{abstract}
There exist a number of models in the literature in which the 
weak interactions are derived from a chiral gauge theory based 
on a larger group than $SU(2)_L\times U(1)_Y$.
Such theories can be constructed so as to be anomaly-free 
and consistent with precision electroweak measurements, 
and may be interpreted as a deconstruction of an extra dimension.
They also provide interesting insights into the issues of flavor and
dynamical electroweak symmetry breaking, and can help to raise
the mass of the Higgs boson in supersymmetric theories.
In this work we show that these theories can also give rise
to baryon and lepton number violating processes, such as nucleon 
decay and spectacular multijet events at colliders, via
the instanton transitions associated with the extended gauge group.
For a particular model based on $SU(2)_1\times SU(2)_2$, 
we find that the $B+L$ violating scattering cross sections are
too small to be observed at the LHC, but that the lower limit on 
the lifetime of the proton implies an upper bound on the 
gauge couplings.

\end{abstract}

\thispagestyle{empty}
\newpage

\setcounter{page}{1}

%%%%%%%%%%%%%%%%%%%%%%%%%%%%%%%%%%%%%%%%%%%%%%%%%%%%%%%
% Introduction
%%%%%%%%%%%%%%%%%%%%%%%%%%%%%%%%%%%%%%%%%%%%%%%%%%%%%%%

\section{Introduction}

Baryon~($B$) and lepton~($L$) number seem to be excellent 
symmetries of Nature, and to date no direct evidence for their
violation has been found.  Even so, it is very likely  
that neither of these charges is exactly conserved.
For one, the Universe contains many more baryons than anti-baryons,
and a necessary ingredient to create such an asymmetry is the violation
of baryon number \cite{Sakharov:1967dj}.  In addition, 
the existence of very small neutrino masses
may also point toward the violation of lepton number.  Such masses
can be naturally generated by the see-saw mechanism   
which typically involves a heavy Majorana neutrino, whose mass
violates lepton number by two units \cite{Gell-Mann:1976pg}.
But perhaps the most compelling reason to expect the violation
of baryon and lepton number is the fact that these charges are not 
even conserved by the Standard Model~(SM) \cite{thooft}.
  
In the SM, both $B$ and $L$ are symmetries of the classical Lagrangian,
but are violated by quantum corrections.  Equivalently, 
the currents corresponding to these would-be symmetries are
anomalous, having non-vanishing divergences.  
However, the only processes that change the value of these charges 
in the SM are \emph{instanton} transitions between degenerate 
$SU(2)_L$ gauge vacua.  Each transition violates both $B$ and
$L$ by $n_g$ units, where $n_g$ is the number of generations.
The rate for these transitions is proportional to a very small
instanton tunnelling factor,
\be
\Gamma_{inst} \propto e^{-16\pi^2/g_L^2} \sim e^{-400},
\label{winst}
\ee
where $g_L$ is the $SU(2)_L$ gauge coupling.  
Because of this enormous suppression, $B$ and $L$ violation
are effectively non-existent in the SM (at zero temperature) 
explaining why neither one has been observed.  Eq.~(\ref{winst}) also 
indicates that the rate would be much larger if the gauge 
coupling $g_L$ were larger.
%, as is sometimes the case 
%in extensions of the SM.

Even though the Standard Model provides an excellent description
of nearly all particle physics interactions seen so far, there is
reason to believe that this model only gives an effective description
of Nature below some ultraviolet cutoff scale.  Above the cutoff,
the SM must be extended to include new physics.  
In many cases the new physics has additional sources
of baryon and lepton number violation.  This can occur through new
perturbative interactions, such as in grand unified theories 
and supersymmetric models with $R$-parity violation.  
The new physics may also violate $B$ and $L$ through 
non-perturbative phenomena,   
%
%Another exciting possibility is that the new physics leads to 
%non-perturbative phenomena that violate $B$ or $L$.
%arises from new non-perturbative effects, 
%
as in models where the electroweak gauge structure is extended beyond
the $SU(2)_L\times U(1)_Y$ group of the SM.  
Depending on the fermion charges under this extended gauge group,
the instanton transitions in such models can violate $B$ and $L$.
Unlike the $SU(2)_L$ rate, however, the instanton rates 
in gauge extended models can be sizeable if the corresponding 
gauge couplings are reasonably large.  This opens the possibility 
of observable baryon and lepton number violating processes within 
these models~\cite{hilltalk}.

In the present work, we examine this possibility for a particular
gauge extension of the SM.  The enlarged electroweak gauge group 
we consider is $SU(2)_1\times SU(2)_2\times U(1)_Y$.  
Under this group, the left-handed fermions of the third generation 
transform as doublets of $SU(2)_1$ and singlets of $SU(2)_2$, 
while the left-handed fermions of the first and second generations are 
doublets of $SU(2)_2$ but singlets of $SU(2)_1$.
The SM electroweak structure is regained by spontaneously breaking 
$SU(2)_1\times SU(2)_2$ down to its diagonal $SU(2)$ subgroup, 
which is identified with the $SU(2)_L$ group of the SM.  
This particular gauge structure arises in several extensions of the SM,
such as Topflavor \cite{Muller:1996dj}, which seeks to motivate 
the hierarchy in the Yukawa couplings, as well as Non-Commuting 
Extended Technicolor \cite{Chivukula:1995gu}, in which the $SU(2)_1$ 
is associated with the ETC gauge group.  Another application 
arises in supersymmetric
theories which increase the tree-level Higgs mass through the $D$-terms
of the extra $SU(2)$ \cite{Batra:2003nj}, as well as 
supersymmetric  models in which baryogenesis is induced by the presence
of strongly interacting Higgsinos and gauginos~\cite{Carena:2004bg}.
Finally, this model is expected to capture,
through dimensional deconstruction \cite{Hill:2000mu},
the low energy physics of an extra dimension with $SU(2)$ in the bulk and
localized fermions \cite{Arkani-Hamed:1999dc}.

When $\su12$ breaks down to its diagonal subgroup,
there are instantonic effects which are not captured by the instantons
of the low energy diagonal $SU(2)_L$ \cite{Csaki:1998vv}.  Thus, we expect
non-perturbative effects in
such theories with extended weak interactions to lead to qualitatively
new effects.  Furthermore, the gauge couplings of the two original
$SU(2)$'s must necessarily be stronger than the diagonal coupling $g_L$,
enhancing the instanton transitions relative to those of $SU(2)_L$.
In several of the examples above, it is further true that 
one of the $SU(2)$ gauge
couplings is considerably larger than the other.  The instanton
transitions of this more strongly-coupled subgroup will then be 
much more frequent than those of the other $SU(2)$.
The observable effects of such instantons are two-fold.
In the context of particle collider experiments such as the LHC,
they can mediate spectacular $B$ and $L$ violating scattering
events.  On the other hand, the violation of baryon and lepton
number also opens the possibility of nucleon decay, and this
puts interesting constraints on these models.
Even though we are focused on a particular gauge extension
of the SM, we also emphasize that we have only made this choice
for concreteness.  For more general gauge extensions of the SM
electroweak sector, we expect that many of our results, as well
as the formalism used to obtain them, to carry over in much 
the same way.

  Previous work along these lines has focused on high energy 
scattering in the SM due to $SU(2)_L$ instantons.  The results of 
Refs.~\cite{Aoyama:1986ej,Ringwald:1989ee,espinosa}
suggest that at very high energies, the sum over high-multiplicity 
exclusive cross sections exponentiates yielding a factor that partially 
cancels the instanton suppression, and producing a potentially observable 
inclusive cross section at future colliders such as the VLHC.  
(See also Refs.~\cite{instascat,Zakharov:1990dj,Shuryak:1991pn,
Farrar:1990vb,Bezrukov:2003er,Mattis:1991bj}.)
However, the approximations made in these calculations generally break 
down at energies below which the instanton suppression is 
significantly reduced.  Instead, in the present work we consider only
exclusive processes due to the instantons of an extended gauge group.  
Our results for collider cross sections will therefore represent 
a conservative lower bound on ($B+L$) violating scattering events 
in these gauge-extended models at the LHC.

  The article is organized as follows.
In Section~\ref{topflavour} we discuss the structure of the 
$SU(2)_1\times SU(2)_2\times U(1)_Y$ gauge extension, 
and describe the bounds on this extension due to precision
electroweak measurements.  Our main results are contained in
Section~\ref{inst} where we outline the formalism used to describe
the instanton transitions within the model, and compute the
effective $B+L$ violating operator generated by $SU(2)_1$
instantons.  In Section~\ref{scatt1} we apply this result 
to calculate the cross section for $B$ and $L$ violating
scattering events at the LHC induced by $SU(2)_1$ instantons.  
Section~\ref{pdecay1} contains an analysis of nucleon 
decay due to $SU(2)_1$ instantons, as well as a discussion
of the constraints implied by this possibility.  The opposite
limit of this scenario, in which the $SU(2)_2$ gauge coupling
is taken to be large, is considered in Section~\ref{light}. 
As in the previous sections, we examine the possibility 
of nucleon decay and $B+L$ violating scattering.
Finally, Section~\ref{conc} is reserved for our conclusions.  
Some of the technical details of our calculations are given 
in the Appendices~\ref{apspin}, \ref{apew}, and~\ref{app2}.

%%%%%%%%%%%%%%%%%%%%%%%%%%%%%%%%%%%%%%%%%%%%%%%%%%%%%%%
%  SU(2)_1 x SU(2)_2, and PEW
%%%%%%%%%%%%%%%%%%%%%%%%%%%%%%%%%%%%%%%%%%%%%%%%%%%%%%%

\section{A Gauge Extension of the Standard Model}
\label{topflavour}

  The gauge extension of the Standard Model that we
consider in the present work is based on the gauge group
$SU(3)_c\times SU(2)_1\times SU(2)_2\times U(1)_Y$.
The $SU(3)_c$ and $U(1)_Y$ subgroups coincide identically
with those of the SM.  On the other hand, the $SU(2)_L$ group 
of the SM is expanded to a larger $\su12$ structure.
While the gauge structure of the SM is extended in this
scenario, the fermion content of the model is identical to the SM.
Under the new $SU(2)_1$ and $SU(2)_2$ groups, the doublets of 
the third generation transform as doublets under $SU(2)_1$ and singlets 
under $SU(2)_2$, while the first and second generation doublets 
are singlets of $SU(2)_1$ and doublets of $SU(2)_2$.  In other words, their
$\su12 \times U(1)_Y$ quantum numbers are
\bea
Q^3 &=& (\mathbf{2}, \mathbf{1})_{1/6}\nnmb\\
L^3 &=& (\mathbf{2}, \mathbf{1})_{-1/2}\nnmb\\
Q^{1,2} &=& (\mathbf{1}, \mathbf{2})_{1/6}\nnmb\\
L^{1,2} &=& (\mathbf{1}, \mathbf{2})_{-1/2}.
\eea

The SM gauge structure is regained by giving a vacuum 
expectation value~(VEV) to a bidoublet scalar, $\Sigma$,
\be
\sig \to \left<\sig\right> = u\;\delta_{i\bar{k}}.
\ee  
Under this breaking, the Standard Model $SU(2)_L$ group emerges
as the unbroken diagonal subgroup of $\su12$.
The corresponding $SU(2)_L$ gauge coupling is
\be
g_L = \frac{g_1g_2}{\sqrt{g_1^2+g_2^2}}.
\ee
This relation implies that when one of the gauge
couplings becomes large, the other one approaches
$g_L$ from above, and thus both $g_1$ and $g_2$ are 
necessarily larger than $g_L$.  The fermion doublets of 
either $SU(2)_1$ or $SU(2)_2$ transform as doublets under $SU(2)_L$.

At a lower scale, $v\simeq 174$~GeV, the remaining
$SU(2)_L\times U(1)_Y$ electroweak symmetry is broken
to $U(1)_{em}$ as in the SM.  This is accomplished by 
giving a VEV to one or more Higgs boson doublets.
We will focus on the case of a single $Y=+1/2$ 
Higgs boson doublet, but our results would be largely 
unchanged if we included instead a $Y=\pm 1/2$ pair of doublets
as in the MSSM.  
We consider two possible representations for the Higgs 
boson under $\su12$.  They are:
\bea
\Phi &=& (\mathbf{2},\;\mathbf{1}
)_{1/2}~~~~~~~\Rightarrow\mbox{\emph{heavy} case},\\
\Phi &=& (\mathbf{1},\;\mathbf{2}
)_{1/2}~~~~~~~\Rightarrow\mbox{\emph{light} case}.\nnmb
\eea
In the first case, which we call the \emph{heavy} case,
the Higgs doublet is charged under $SU(2)_1$ but not under $SU(2)_2$.
The opposite is true for the \emph{light} case.
These two possibilities are very similar with regards to instantons,
but differ significantly when it comes to the experimental
constraints on the model.  We will consider them both.
In Appendix~\ref{apew} we tabulate some important results 
concerning the gauge bosons, their masses, and their 
couplings to fermions.

\subsection{Precision Electroweak Constraints}

  The most important experimental constraints on this gauge
extended model come from precision electroweak measurements made
at LEP, the Tevatron, and the SLC.  Due to the enlarged gauge structure,
the model has additional heavy gauge bosons, a ${Z^0}'$ 
and a ${W^{\pm}}'$,
and modified relations between the Lagrangian parameters and 
the electroweak observables.  The gauge boson mass matrices
and the shifts in the electroweak observables are listed
in Appendices~\ref{apew} and ~\ref{app2}.
Because of these changes, the precision electroweak 
data imposes strong constraints on the model, and on the 
$\su12$ symmetry breaking scale $u$ in particular.
The precise constraints are different for the heavy 
and light cases described above. 

   The Lagrangian-level parameters of the electroweak sector 
of the model can be taken to be $\{g_1,\,g_2,\,g_y,\,v,\,u\}$.  
We find it more convenient to use the equivalent
set $\{g_L,\,v,\,\sin\theta,\,\sin\varphi,\,\delta\}$, where
\be
\begin{array}{cccccc}
g_L &=& \frac{g_1g_2}{\sqrt{g_1^2+g_2^2}},&~~~~~
\sin\theta &=& \frac{g_y}{\sqrt{g_y^2+g_L^2}},\\
\sin\varphi &=& \frac{g_2}{\sqrt{g_1^2+g_2^2}},&~~~~~
\delta &=&\frac{v^2}{2\,u^2}.
\end{array}
\ee
All (tree-level) electroweak observables can be expressed
in terms of these.  In our analysis, we specify the values
of $\delta$ and $\sin\varphi$, and use the measured values
of $M_Z(M_Z)$, $\alpha(M_Z)$, and $G_F$ (extracted from 
the muon decay rate) to fix the rest. 
We use the values~\cite{Eidelman:2004wy},
\bea
\alpha^{-1}(M_z) &=& 127.918\nonumber\\
G_F &=& 1.16637\times\,10^{-5}\;\mbox{GeV}^{-2}\nonumber\\
M_Z(M_Z) &=& 91.1876 \;\mbox{GeV} ~. \nonumber
\eea

\begin{figure}[t]
\centerline{
\hspace*{-1cm}
        \includegraphics[width=0.5\textwidth]{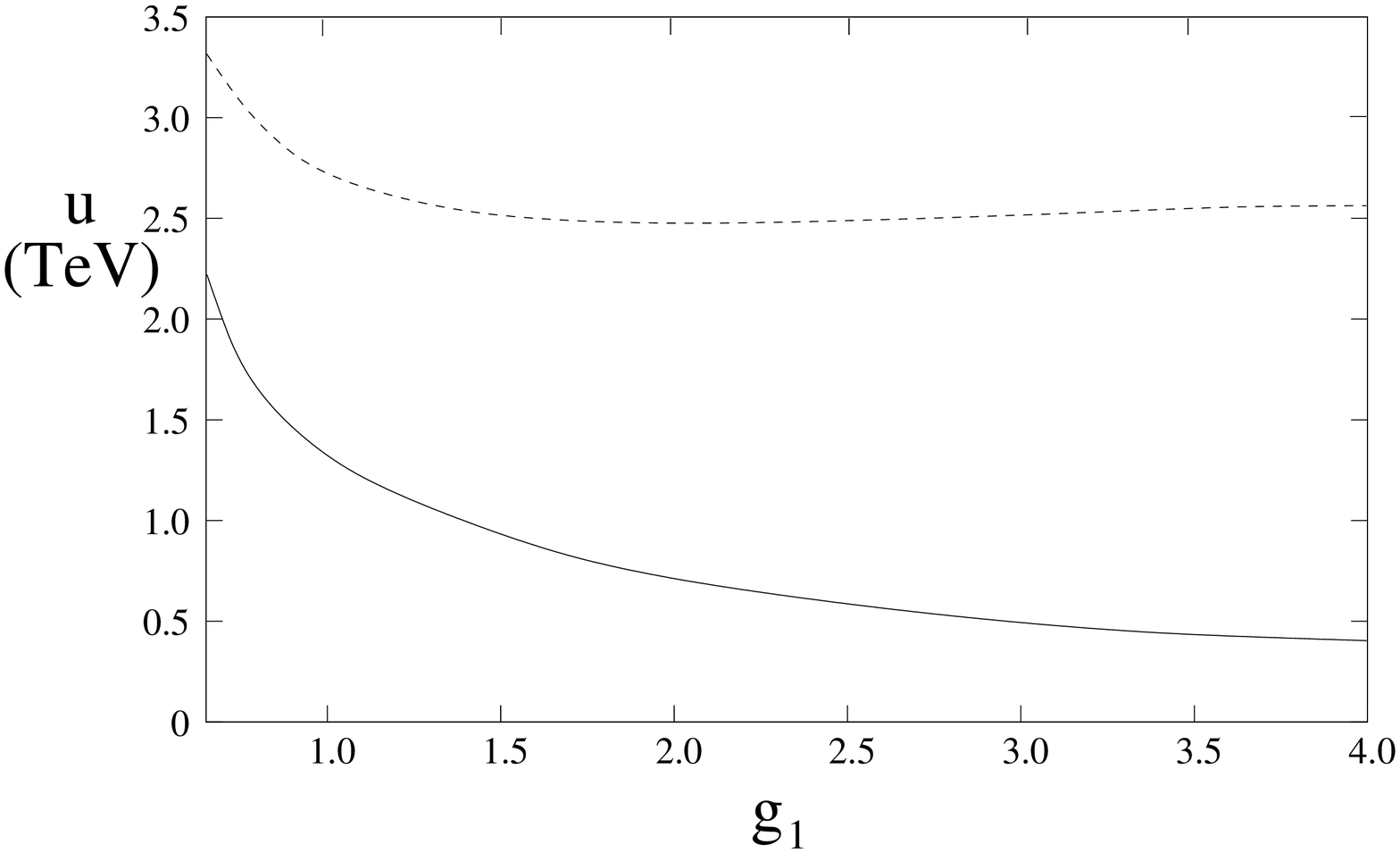}
        \includegraphics[width=0.5\textwidth]{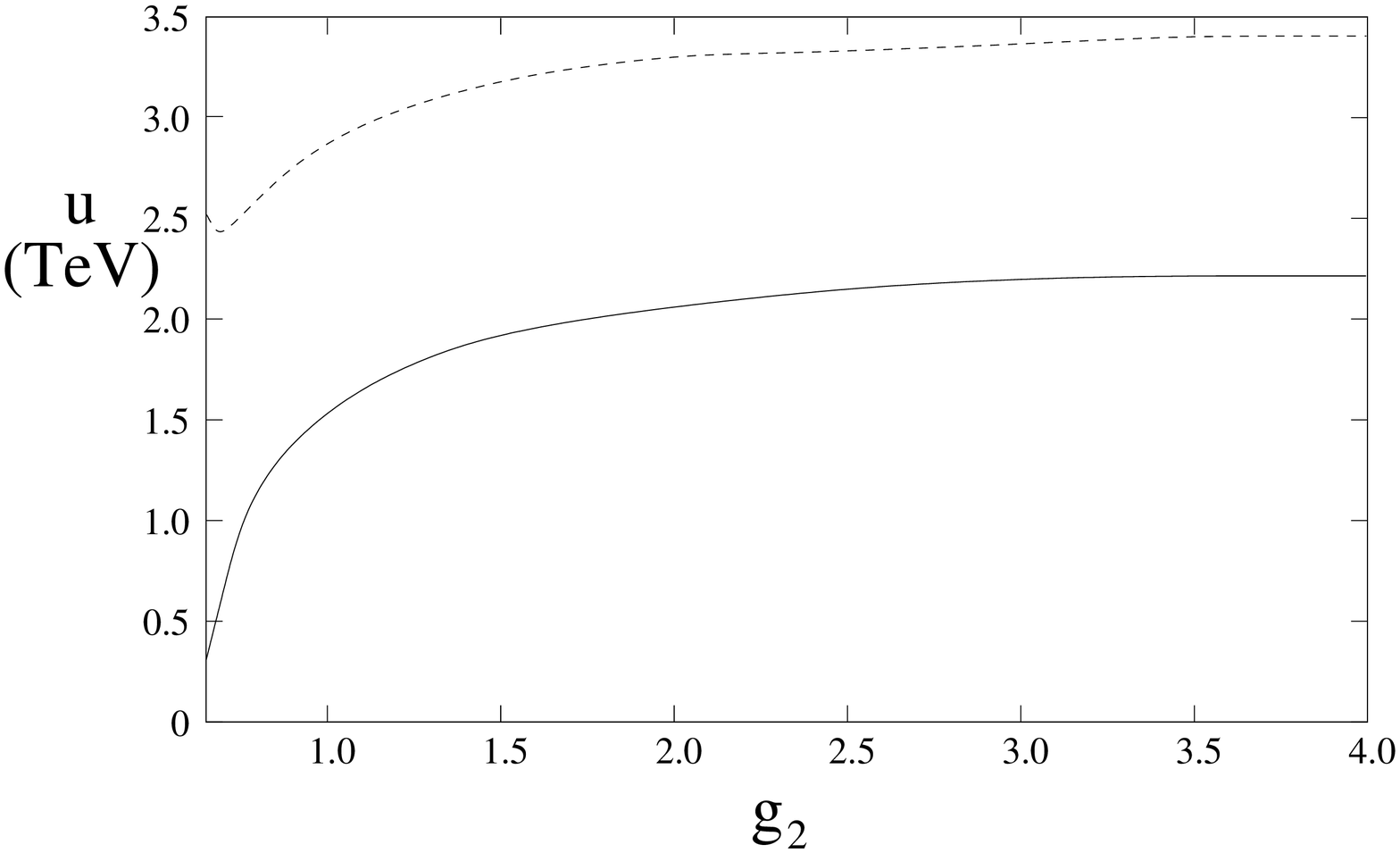}}
        \caption{$95\%~c.l.$ exclusion contour as a function
of $g_1$ and $g_2$.  The allowed region lies above the solid curve
~(\emph{light} case) or the dashed curve~(\emph{heavy} case).}
\label{ewfitg1}        
\end{figure}

Having fixed and specified the electroweak parameters,
we may calculate the shifts in the electroweak observables
due to the extended gauge structure.  For example, the shift
in the $W$ mass compared to the SM in the \emph{light} case is
\be
M_W = (M_W)_{SM}(1 + 0.219\,\sin^4\varphi\,\delta).
\ee
Here, $(M_W)_{SM}$ should properly be the tree-level expression of the SM.
However, if we work to first order in both the loop corrections
and the small parameter $\delta$, it is consistent to use the
one-loop value of $(M_W)_{SM}$ in this expression.
The shifts in other important electroweak observables are 
tabulated in Appendix~\ref{app2}.\footnote{
$\Gamma_{inv}$ and $\Gamma_{e,\mu}$ were not included
in the analysis since the first is not directly observable, and the second
is not an independent quantity once $\Gamma_Z,\:\Gamma_{had}$, 
and $R_{e,\mu,\tau}$ have been used.}
A Higgs boson mass of $m_h = 115$~GeV and a top quark 
mass of $m_t = 177$~GeV were used to obtain the SM inputs.  
For each parameter set we compute the effective reduced $\chi^2$:
\be
\chi^2 = \sum_{i=1}^{N} \frac{(\mathcal{O}_i
-\mathcal{O}_i^{exp})^2}{\sigma_i^2},
\ee
where $\mathcal{O}_i$ is the value of the $i$-th observable in 
the model, $\mathcal{O}_i^{exp}$ is the measured value of 
this observable, and $\sigma_i$ is its experimental 
uncertainty.
We demand that $\chi^2/N < 1.6$, which corresponds (roughly) 
to the $95\%~c.l.$ exclusion contour for $N=20$ degrees of freedom.  
(By comparison, the best fit to the SM, for the observables 
considered, has a $\chi^2/N = 1.03$.)  
The exclusion contours are shown in Fig.\ \ref{ewfitg1}.  
Observe that in the light case, the bounds
on $u$ become very weak for large values of $g_1$ because only
the third generation sector is affected by the strong interactions
(resulting in no large corrections to $G_F$ extracted from muon decay),
and the mixing between the light and heavy gauge bosons induced
by the standard Higgs VEV becomes smaller for larger values
of $g_1$.

As discussed above, the above bounds on $u$
were obtained for a Higgs mass close to the 
present experimental bound. These bounds may not be lowered in
any significant way by raising the Higgs mass. 
In the light case, raising the Higgs 
mass up to values close to 200~GeV produces very small 
variations in the bound on $u$.  In the heavy case, the bound 
on $u$ increases with the Higgs mass. For instance, for a Higgs boson
mass of about 150~GeV, the lower bound on $u$ increases by about 
500 GeV for all values of $g_1 > 1.5$.

%%%%%%%%%%%%%%%%%%%%%%%%%%%%%%%%%%%%%%%%%%%%%%%%%%%%%%%
%  Instanton Stuff, in General
%%%%%%%%%%%%%%%%%%%%%%%%%%%%%%%%%%%%%%%%%%%%%%%%%%%%%%%

\section{Instanton Induced Operators}
\label{inst}

In this section, we derive effective operators which describe the 
instanton-induced interactions at low energies.  We begin with some 
general features of instantons in broken gauge theories, and then specialize
to the case of $\su12$.  It is well-known that
non-Abelian gauge theories have many physically 
distinct vacua separated by energy barriers of finite height.
As a result, it is possible for a system prepared in one vacuum
state to pass to another by tunnelling.  The gauge field configurations
that describe this tunnelling are called instantons.  
As we shall see, if there are fermions charged under the gauge group,
each instanton transition is accompanied by the production of fermions.  
For $SU(2)_L$ instantons in the SM,
this is the source of $B$ and $L$ violation.

In a pure non-Abelian gauge theory, instanton configurations 
are solutions of the Euclidean space equations of motion
with finite Euclidean action.
A given instanton solution is characterized by its spacetime 
location, $x^{\mu}_0$, its Euclidean space radius, $\rho$, 
and its orientation in the global gauge group space, $U$.
The instanton transition amplitude is computed by making
a semiclassical expansion of the corresponding functional 
integral about the instanton solution, working to quadratic 
order in the fluctuations about this solution.  
This procedure generates a factor of $e^{-S_{inst}} = 
e^{-8\pi^2/g^2}$ from the classical solution, as well as a 
functional determinant from the fluctuations~\cite{thooft}.

The situation becomes more complicated if the gauge theory is 
spontaneously broken by the expectation value of one or more scalar fields.
In this case, exact solutions to the combined gauge/Higgs 
Euclidean space equations of motion are not known. 
Nevertheless, it is possible to obtain approximate solutions 
for a fixed instanton size, $\rho$, as expansions in 
$\rho\left<\phi\right>$, where $\left<\phi\right>$ is the 
symmetry breaking VEV~\cite{affleck}.  For a given $\rho$, 
the contribution of the Higgs field to the 
Euclidean action is~\cite{espinosa,affleck}
\be
S_{Higgs} = 2\pi^2\rho^2\left<\phi\right>^2
+ \mathcal{O}(\lambda\,\rho^4\left<\phi\right>^4),\label{shiggs}
\ee
where $\lambda$ denotes a quartic coupling for the scalars.
The full transition amplitude is given by the fixed--$\rho$
amplitude integrated over instanton size.  Since the integrand 
is proportional to $e^{-S_{Higgs}}$, this integral is cut off 
at $\rho\left<\phi\right> \sim 1/\sqrt{2\pi^2}$ justifying the
expansion in this parameter.
The leading contribution from the Higgs field to the action, 
Eq.~(\ref{shiggs}), comes from the kinetic term since 
interactions are higher order in $\rho\left<\phi\right>$.  
Thus, if there are several scalar 
multiplets which develop VEV's, the leading contribution to the 
action will be the sum of the individual contributions, 
each with the form of Eq.~(\ref{shiggs}).
Note, however, that it is only possible to neglect 
the interaction term in $S_{Higgs}$ if the scalar quartic coupling 
is not too large, $\lambda \ll 2\pi^2$, which we will assume in
the present work.  On the other hand, for $\lambda \to \infty$ 
the transition amplitude, being proportional 
to $e^{-S_{Higgs}}$, vanishes~\cite{D'Hoker:1984jq}.  In this limit, 
the symmetry breaking sector may be represented by a non-linear 
sigma model, and the vanishing of the transition amplitude can be 
explained by the existence of a conserved topological 
current~\cite{chill}.  The transition between the small and 
large $\lambda$ regimes is an interesting question, but requires
a precise specification of the symmetry breaking sector, and is 
outside the scope of the present work.

If the theory also has fermions that are charged under the 
gauge group, this picture of vacuum tunnelling is changed in 
an important way.  While the fermions do not modify the classical
instanton solution (at lowest order), the functional integral over 
the quantum fluctuations now includes an integration over the 
Grassmann-valued fermion fields.  
The integral vanishes unless it is saturated by fermions 
from the integrand.  For a trivial (zero instanton) background, 
this leads to a non-zero fermion determinant.  
However, in an instanton background 
there exist fermionic fluctuations which do not contribute 
to the action at quadratic order.\footnote{Equivalently, the fermion
bilinear operator has one or more zero eigenvalues in the 
instanton background.}  These fermion \emph{zero modes} 
are nonetheless part of the functional integration,
and the amplitude vanishes.  In general, for each fermion representation
$r$, there are $2\,T(r)$ fermion and no anti-fermion zero modes in
a one-instanton background~\cite{terning}.

While the vacuum transition amplitude vanishes if there are fermions 
coupled to the gauge group, a non-zero result is obtained 
if an appropriate number of fermion fields, one for each zero mode, 
are inserted into the functional integral.    
The instanton transitions are therefore accompanied by the 
production of fermions.  For the case of $SU(2)_L$ instantons,
there are $4\,n_g$ fermion doublets, three quark doublets and 
one lepton doublet for each generation, and therefore $4\,n_g$
zero modes.  The corresponding transition violates both
$B$ and $L$ by $n_g$ units.  For $SU(2)_1$ and $SU(2)_2$
instantons in the gauge-extended model described in the previous section, 
the result is the same except now $n_g =1$ or $2$.
Thus, the instantons in all three cases violate $B+L$.

\subsection{Instanton Green's Functions}

  In this section we describe the calculation of 
instanton-induced fermion Green's functions for a general
$SU(2)$ gauge theory with $n_f$ Weyl fermion
doublets, an arbitrary number of fermion singlets,
and $n_s$ complex scalar doublets.  
There are $n_f$ fermion zero modes in this case, and
the resulting Green's function will involve one of each of
the fermion doublets.
The presentation here follows the discussions
of Ringwald~\cite{Ringwald:1989ee} 
and Espinosa~\cite{espinosa}.  Both of these, 
in turn, rely heavily on the results of 't~Hooft~\cite{thooft}. 

We wish to calculate the Green's function
\be
G(x_1,\ldots,z_m) = 
\left<\prod_{i=1}^{n_f}\psi_i(x_i)\;\prod_{j=1}^nA_{\mu_j}^{a_j}(y_j)\;
\prod_{k=1}^m H(z_k)\right>_{1-inst.},
\ee
where the $\psi$ are fermions, the $A$ are gauge fields, and the
$H$ are (shifted) scalar fields ($\Phi = \left<\phi\right> + H$). 
%This Green function can be computed in the semiclassical approximation
%by expanding the corresponding path integral about the classical
%(Euclidean space) instanton solution and working to quadratic order
%in the residue.  

  Following~\cite{thooft,espinosa}, the combined gauge boson and 
Higgs boson instanton solution is 
\bea
\label{eqn:bg1}
A_{\mu} &=& x_{\nu}\,\Lambda_{\mu\nu}\;\mathscr{A}(x^2),\\
\Phi(x) &=& \phi(x^2)\;\bar{h},\nonumber
\eea
with $\bar{h} = (0,1)^t$, and $\Lambda_{\mu\nu} = 
U\bar{\tau}_{\mu\nu}U^{\dagger}$, where $\bar{\tau}_{\mu\nu}$
is the matrix $\bar{\sigma}_{\mu\nu}$ acting in the $SU(2)$
space and $U$ is an $SU(2)$ matrix describing the instanton 
orientation.  Their explicit forms are listed below and 
in Appendix~\ref{apspin}.
The functions $\mathscr{A}$ and $\phi$ have asymptotic expressions
valid at large and small distances, respectively:
\bea
\label{eqn:bg2}
\mathscr{A}(x^2) &=& \left\{\begin{array}{lr}
\frac{1}{g}\,\frac{2\rho^2}{x^2(x^2+\rho^2)},&x\ll\rho
\vspace{0.3cm}\\
\frac{1}{g}\,\rho^2M_W^2\frac{K_2(M_Wx)}{x^2},&x\gg\rho
\end{array}
\vspace{0.3cm}\right.\\
\phi(x^2) &=& \left\{\begin{array}{lr}
\left(\frac{x^2}{x^2+\rho^2}\right)^{1/2}\avg{\phi},&x\ll\rho
\vspace{0.3cm}\\
\avg{\phi} - \frac{1}{2}\rho^2m_h\avg{\phi}\frac{K_1(m_hx)}{x},&x\gg\rho.
\end{array}\right. \nonumber
\eea
The long distance forms are leading term expansions in 
$\rho \langle \phi \rangle$.
These functions correspond to the \emph{singular gauge}, which has 
the useful property that the gauge fields go to zero at 
Euclidean infinity.

  Using these solutions, the semiclassical approximation 
to the functional integral gives~\cite{espinosa}
\bea
G(x_1,\ldots,z_n) &=&\\
&&\phantom{\!\!\!\!\!\!\!\!\!\!\!\!\!\!\!\!\!\!\!\!\!\!
\!\!\!\!\!\!\!\!\!\!\!\!\!\!\!\!\!\!\!\!\!\!\!\!}
\int\!d^4x_0\!\int\!d\rho\!\int\!(dU/8\pi^2)\;
\tilde{F}(\rho,\left<\phi\right>;\mu)\;
e^{-S_E[A_{cl},\Phi_{cl}]}\;\prod_{i=1}^{n_f}\psi_{0_i}(x_i-x_0)
\prod_{j=1}^nA_{cl}(y_j-x_0)\prod_{k=1}^{m}H_{cl}(z_k-x_0),\nnmb
\eea
where $A_{cl}$ and $\Phi_{cl}$ are the classical instanton solutions
given above ($H_{cl}=\Phi_{cl}-\left<\phi\right>$), 
and $\psi_{0_i}$ is the $i$-th fermion zero mode 
in the instanton background.
The integrals over the instanton size $\rho$, location $x_0$,
and group orientation $U$ correspond to collective coordinates 
for the functional integrations over the zero modes of the gauge 
field fluctuations.  Finally, $\tilde{F}(\rho,\left<\phi\right>;\mu)$ 
is a product of functional
determinants for the non-zero vector, scalar, and fermion modes,
along with the Jacobian factors from converting to collective coordinates.  

  For the approximate instanton solution in the combined gauge/Higgs
system, the Euclidean action at leading order in $\rho\left<\phi\right>$
is given by
\be
S_E[A_{cl},\Phi_{cl}] = \frac{8\pi^2}{g(\mu)^2} 
 + 2\pi^2\rho^2\,\vv^2
%\sum_{k=1}^{n_s}\left<\phi_k\right>^2
\label{action}
\ee
where
\be
\vv^2 = \sum_{k=1}^{n_s}\left<\phi_k\right>^2.
\label{vv}
\ee
The factors comprising $\tilde{F}(\rho,\left<\phi\right>;\mu)$ were
calculated in~\cite{thooft},
\be
\tilde{F}(\rho,\left<\phi\right>;\mu) = C\,g^{-8}(\rho\mu)^{b_0}
\rho^{n_f/2-5}
\ee
where 
\be
b_0 = \frac{22}{3} - \frac{1}{3}\,n_f - \frac{1}{6}\,n_s
\ee
is the one loop beta-function coefficient,
and the constant $C$ is given by
\be
C = 2^{10}\pi^6\,\exp\left[ \left(8-\frac{1}{2}n_f\right)\,\xi_a 
- \left(\frac{2}{3} - \frac{1}{6}n_f + \frac{1}{6}n_s\right)\xi_b 
- \alpha(1) + (n_f-n_s)\alpha\left(\frac{1}{2}\right)\right].
\label{prefactor}
\ee
Here, $(\xi_a, \xi_b) = (0,-5/12)$ for $g(\mu)$ defined in the
$\overline{MS}$ scheme, and $\alpha(1)$ and $\alpha(1/2)$ are
numerical constants with the approximate values
\be
\alpha(1) \simeq 0.443,\phantom{AAAAAAA}
\alpha(1/2) \simeq 0.146.
\ee
The additional factors of $\rho$ are inserted to get the 
dimensions right.\footnote{Note that 
$[\psi_0(x)] = M^2$. Also, the expression for $C$ differs from 
the corresponding expression given by Espinosa~\cite{espinosa} 
by a factor of $(8\pi^2)^{n_f/2}$.  For comparison, 
in Ref.~\cite{thooft}, this factor arises from the normalization 
of an effective operator describing the instanton coupling to fermions.  
Here, no such operator has been inserted so this factor 
is redundant. There is also an additional factor of ${1}/{8\pi^2}$ 
in the measure of the $U$ integral since we are explicitly
keeping the integral over global gauge rotations.}
% \\{\bf Get rid of footnote?}\\
Note also that the combination
$\mu^{b_0}e^{-8\pi^2/g(\mu)^2}$ is RG-invariant at one-loop order.

  Upon Fourier transforming, the $d^4x_0$ integral generates
a total momentum conserving delta function.  
The momentum space Green's function, cancelling off a 
$(2\pi)^4\,\delta^{(4)}(\sum p_i)$ factor, is therefore
\be
\tilde{G}(\{p\},\{q\},\{k\}) = 
\int\!(dU/8\pi^2)\!\int\!d\rho\;\tilde{F}(\rho,\left<\phi\right>;\mu)\;
e^{-S_E[A_{cl},\Phi_{cl}]}\;\prod_{i=1}^{n_f}\tilde{\psi}_{0_i}(p_i)
\prod_{j=1}^n\tilde{A}_{cl}(q_j)\prod_{k=1}^{m}\tilde{H}_{cl}(k_k),
\label{igreen}
\ee
where $\tilde{\psi}_0$, $\tilde{H}_{cl}$, and $\tilde{A}_{cl}$ 
denote the Fourier transforms.

\subsection{Fermion Zero Modes}

  To proceed, we need explicit expressions for the fermion zero 
modes, and for this, we must specify the couplings between
the fermions and the scalars.  We will focus on the gauge
extended model described in Section~\ref{topflavour}, and look
at the instantons of the $SU(2)_1$ group that couples to 
the third generation and the Higgs doublet (\emph{heavy} case).  
These solutions are identical to those for $SU(2)_L$ instantons obtained 
in Ref.~\cite{espinosa}, and also carry over directly for $SU(2)_2$ 
instantons in the \emph{light case}.
Unlike Ref.~\cite{espinosa}, however, we use a slightly different 
set of Euclidean space spinor conventions, and because of this, 
our results are somewhat different in appearance.
These conventions are listed in Appendix~\ref{apspin}.

  In Euclidean space, unlike Minkowski space, the two spinor
representations of $SO(4)$ are not related by complex conjugation.
Instead, the two $SO(4)$ representations, which we label by $A$ 
and $B$, are related to those of $SO(1,3)$ via the correspondence
\be
\begin{array}{cccccc}
\psi_R &\leftrightarrow& \psi_A,\phantom{AAA}
&\psi_L &\leftrightarrow& \psi_B,\\
\psi_R^{\dagger} &\leftrightarrow& \psi_B^{\dagger},\phantom{AAA}
&\psi_L^{\dagger} &\leftrightarrow& \psi_A^{\dagger}.
\end{array}
\ee
Using this relation, the equations satisfied by the 
quark zero modes are
\bea
\label{eqn:zerom}
0 &=& \bar{\sigma}_{\mu}{D}_{\mu}\,Q_B, 
-i\lambda_u\epsilon\,\Phi^*_{cl}\,u_A -i\lambda_d\,\Phi_{cl}\,d_A,
\nonumber\\
0 &=& \sigma_{\mu}\del_{\mu}u_A -i\lambda_u\,\Phi^t_{cl}\epsilon\,Q_B,\\
0&=&\sigma_{\mu}\del_{\mu}d_A +i\lambda_d\,\Phi^{\dagger}_{cl}\,Q_B,
\nonumber
\eea
where $D_{\mu} = \del_{\mu} -igA_{cl\,\mu}$, $\epsilon = i\sigma^2$,
and the $\lambda_i$ are Yukawa interactions.
$Q_B$ corresponds to the left-handed quark doublet, 
$u_A$ and $d_A$ are the Euclidean forms of the right-handed singlets,
and $\Phi_{cl}$ and $A_{cl}$ denote the classical instanton solutions
given above.  The equations for the lepton zero modes have 
the same form.

  To solve Eqs.~(\ref{eqn:zerom}), we insert the background 
solutions from Eqs.~(\ref{eqn:bg1}) and (\ref{eqn:bg2}),
and use the ansatz
\be
\psi_B = x_{\mu}\sigma_{\mu}\,\varphi(x^2),\phantom{AAA}
\psi_A = \psi_A(x^2),
\ee
where $\varphi(x^2)$, like $\psi$, denotes a two-component fermion.
The long-distance equations can be simplified by making use
of $\mathscr{A}(x^2)\to 0$ and $\phi(x^2)\to \avg{\phi}$
for $|x^2|\gg\rho^2$.  The solutions in this case are
\bea
\label{eqn:zmlong}
\psi_B &=& \frac{1}{2\pi}\rho\,m^2\frac{K_2(m\,x)}{x^2}
x_{\mu}\sigma_{\mu}\;U^{\dagger}\chi,\nonumber\\
\psi_A &=& \frac{1}{2\pi}\rho\,m^2\frac{K_1(m\,x)}{x}\,
U^{\dagger}\chi,
\eea
where $m$ is the fermion mass, and $\chi$ represents a two-component
spinor equal to $\chi = {-1\choose \:0}$ for $\psi = d$ 
and $\chi = {0 \choose 1}$ for $\psi = u$.  
At short distances, $|x^2|\ll\rho^2$, the solutions at
leading order in $\rho\avg{\phi}$ are given by
\bea
\label{eqn:zmshort}
\psi_B &=& \frac{1}{\pi}\frac{\rho}{(x^2)^{1/2}(x^2+\rho^2)^{3/2}}
x_{\mu}\sigma_{\mu}\;U^{\dagger}\chi\nonumber\\ 
\psi_A &=& \frac{i}{2\pi}\rho\,m\frac{1}{x^2+\rho^2}\,U^{\dagger}\chi.
\eea

  To obtain the low-energy effective operators generated by 
the instanton, we will need the Fourier transforms
of the long distance zero-mode solutions given by Eq.~(\ref{eqn:zmlong}).
The following (Euclidean space) identities are useful for this:
\bea
\int d^4x\;e^{-ip \cdot x} f(x^2) &=& \frac{4\pi^2}{p}
\int_0^{\infty}dr\;J_1(pr)\,r^2\,f(r^2),\nonumber\\
\int d^4x\;e^{-ip \cdot x}\,x_{\mu}\,f(x^2) &=& 
-{4\pi^2\,i}\,\frac{p_{\mu}}{p^2}
\int_0^{\infty}dr\;J_2(pr)\,r^3\,f(r^2),
\eea
and
\be
\int_0^{\infty}dr\;J_2(pr)\,r\,K_n(m\,r) =
\frac{p^n}{m^n(p^2+m^2)},
\ee
where $p = (|p_{\mu}p_{\mu}|)^{1/2}$.
Applying these identities to the previous result, we find
\bea
\tilde{\psi}_B(p) &=& -2\pi\,i\,\rho
\left(\frac{p_{\mu}\sigma_{\mu}}{p^2+m^2}\right)\,U^{\dagger}\chi,\\
\tilde{\psi}_A(p) &=& -2\pi\,i\,\rho
\left(\frac{m}{p^2+m^2}\right)\,U^{\dagger}\chi.\nonumber
\eea
In the above, the tildes denote Fourier transformed functions.
Since the fermions are massive, it helps to assemble them into
a Dirac fermion and revert to Minkowski space.  The result is
\bea
%\label{eqn:zm}
\label{fzero}
\tilde{\Psi}(p) &=& 2\pi\,i\,\frac{\rho}{p^2-m^2}
\left(p_{\mu}\gamma^{\mu}\,P_R + m\,P_R\right)
{U^{\dagger}\chi \choose U^{\dagger}\chi}\\
&=&\frac{i(p_{\mu}\gamma^{\mu}+m)}{p^2-m^2}\left[2\pi\rho
{0 \choose U^{\dagger}\chi}\right].\nonumber
\eea
As before, $\chi = {-1\choose \:0}$ for $\Psi = d,\;e$, 
and $\chi = {0 \choose 1}$ for $\Psi = u,\,\nu$.

  The same Bessel function and Fourier transform 
identities can be applied to obtain the long distance forms
of the classical gauge and Higgs boson solutions given in
Eqs.\ (\ref{eqn:bg1}), (\ref{eqn:bg2}).  
Reverting to Minkowski space, they are~\cite{espinosa}
\bea
\tilde{A}_{cl_{\mu}}(p) &=& \frac{i}{g}\;\frac{4\pi^2\,\rho^2}
{p^2-m_A^2}\,U\bar{\tau}_{\mu\nu}U^{\dagger}p_{\nu},
\nonumber\vspace{0.2cm}\\
\tilde{H}_{cl}(p)&=& -\frac{2\pi^2\rho^2\left<\phi\right>}{p^2-m_H^2}.
\label{residues}
\eea

\subsection{Instanton Amplitudes}

  With the explicit zero-mode expressions in hand, we may now
construct amplitudes for instanton-induced processes.
Applying the LSZ procedure~\cite{LSZ} to Eq.~(\ref{igreen}), and 
using Eqs.\ (\ref{fzero}) and (\ref{residues}),
the one-instanton amplitude for a process involving $n_g$ 
SM generations ($n_f = 4\,n_g$), 
$n$ gauge bosons, and $m$ scalars is given by~\cite{espinosa}
\bea
\mathcal{A} &=& \frac{C}{g^{8+n}}\mu^{b_0}e^{-8\pi^2/g^2(\mu)}
(4\pi^2)^n(2\pi^2)^m(2\pi)^{4n_g}V^m\cdot\nonumber\\
&&\phantom{AAA}\cdot\left(\int_0^{\infty}d\rho\;\rho^{6n_g-5+2m+2n+b_0}
e^{-2\pi^2\vv^2\rho^2}\right)\;\int({dU}/{8\pi^2})\;h(U),
\eea
where
\be
h(U) = \prod_{i=1}^{4n_g}\bar{\eta}_i(p)
{0 \choose U^{\dagger}\chi}
\prod_{j=1}^{n}\epsilon_{\mu}^{(j)}(q_j)q_{j\nu}
tr(U\bar{\tau}_{\mu\nu}U^{\dagger}\mathscr{P}_j),
\ee
where $\eta_i(p) = u_i(p)$ or $v_i(p)$ is the external state
polarization spinor, and $\mathscr{P}$ projects onto the appropriate 
gauge boson mass eigenstate.  

  The $\rho$ integral is straightforward,
and gives the factor 
\be
\frac{1}{2}\left(\frac{1}{2\pi^2\vv^2}\right)^{3n_g-2+m+n+b_0/2}
\Gamma(3n_g-2+m+n+b_0/2).
\ee
The resulting amplitude (up to an overall phase) is therefore 
\bea
\mathcal{A} &=& \frac{C}{g^{8+n}}e^{-8\pi^2/g^2(\mu)}
\left(\frac{1}{4\pi^2}\right)^{n_g-2+b_0/2}\,
2^{3n_g-3+n+b_0/2}\,\Gamma(3n_g-2+m+n+b_0/2)\,
\left(\frac{\mu}{\vv}\right)^{b_0}\cdot\nonumber\\
&&\phantom{AAAAA}\cdot\left(\frac{1}{\vv^2}\right)^{3n_g-2+m/2+n}
\;\int({dU}/{8\pi^2})\;h(U).
\label{eqn:coeffa}
\eea
In these expressions $\vv^2$ is the orthogonal sum of 
the scalar VEV's, Eq.~(\ref{vv}).  For the case of $SU(2)_1$ or $SU(2)_2$
instantons, the bidoublet field $\Sigma$ transforms as a 
pair of doublets under of these groups, each of which 
develops a VEV equal to $u\sim \mbox{TeV}$.  Thus
\be
\vv^2 = v^2 + 2\,u^2\simeq 2\,u^2\phantom{AAAAA}SU(2)_1~\mbox{or}~SU(2)_2~ 
\mbox{instantons}.
\ee
The VEV of the $\Sigma$ field is  along a singlet
component of $SU(2)_L$, and therefore
\be
\vv^2 = v^2\phantom{AAAAA}SU(2)_L ~\mbox{instantons}.
\ee

\subsection{Instanton Effective Operators for $SU(2)_1$}

  For the remainder of this section, we will focus on the 
situation in which $g_1\gg g_2$, where the instantons
of the $SU(2)_1$ gauge theory become unsupressed.
We would like to represent the amplitude for these instantons,
Eq.~(\ref{eqn:coeffa}), by an effective operator valid below
the $SU(2)_1$-breaking scale.
The amplitude found above corresponds to the Green's function 
$\left<q^1(p_1)\,q^2(p_2)\,q^3(p_3)\,l(p_4)\right>$,
and consists of one zero mode wavefunction for each fermion, 
a numerical prefactor, integrations over the instanton size 
$\rho$ and orientation $U$, and an overall factor of 
$(2\pi)^4\delta^{(4)}(p_1+p_2+p_3+p_4)$ from the integration 
over instanton location.  Since only the total
momentum is conserved, we will be able to represent the 
large-distance instanton effects by a local operator.
Note that since we will use the long-distance expressions of the 
fermion zero modes, which lose validity at energy scales of order 
$E_u \simeq \sqrt{2} \pi u$, the derived effective theory will 
also lose validity at energies larger than  $E_u$.   

  For the task at hand, it is more convenient to look at
the operator generated by an \emph{anti-instanton}.  
In this case, the non-vanishing Green's function is 
$\left<\bar{q}^1\bar{q}^2\bar{q}^3\bar{l}\right>$.
After applying the LSZ procedure, each of
the four fermion zero modes generates a factor of the form 
\be
2\pi\,\rho\left(\chi^{\dagger}U,\;0\right)\eta(p),
\ee
where $\eta(p) =u(p)$ or $v(p)$ is the external-state polarization spinor.  
The resulting amplitude is therefore proportional to
\be
(2\pi\,\rho)^4\,\int dU\,
\left(\chi_1^{\dagger}U,\;0\right)\eta_1(p_1)\cdot
\left(\chi_2^{\dagger}U,\;0\right)\eta_2(p_2)\cdot
\left(\chi_3^{\dagger}U,\;0\right)\eta_3(p_3)\cdot
\left(\chi_4^{\dagger}U,\;0\right)\eta_4(p_4),
%[\bar{u}(p_2)\,\omega_2]\;
%[\bar{u}(p_3)\,\omega_3]\;[\bar{u}(p_4)\,\omega_4]
\ee
with $\chi_i = {0\choose 1}$ for $u$ or $\nu$,
and $\chi_i = {-1 \choose 0}$ for $d$ or $e$.\footnote{
In this section we will denote 
$u\sim t$, $d\sim b$, $\nu\sim \nu_{\tau}$, $e\sim \tau$.}

  To perform the integration over instanton orientation $U$,
we make use of the fact that, as a manifold, $SU(2)$ is equivalent 
to $S^3$.  This equivalence allows us to parametrize an arbitrary 
$SU(2)$ element as  
\bea
U &=& e^{i\,\alpha\,\hat{n}\cdot\vec{\sigma}}\nonumber\\ 
&=& \cos(\alpha) + i\,(\hat{n}\cdot\sigma)\sin(\alpha),\\
&=&\left(\begin{array}{cc}
\cos\alpha + i\sin\alpha\,\cos\theta&
i\,e^{-i\phi}\,\sin\alpha\,\sin\theta\\
i\,e^{i\phi}\,\sin\alpha\,\sin\theta&
\cos\alpha - i\sin\alpha\,\cos\theta
\end{array}\right)\nonumber
\eea
where $\hat{n} = (\sin\theta\cos\phi,\sin\theta\sin\phi,\cos\theta)$ 
is a unit 3-vector.  

The coordinate ranges are
\be
\alpha\in[0,\,\pi],\phantom{AAA}\theta\in[0,\,\pi],
\phantom{AAA}\phi\in[0,\,2\pi],
\ee
and the integration measure is 
\be
\int\,dU = \int_0^{\pi}\!d\alpha\sin^2\alpha\int_{-1}^{1}
\!d(\cos\theta)\int_0^{2\pi}\!d\phi.
\ee

  The Green's functions $\left<\bar{q}^1\bar{q}^2
\bar{q}^3\bar{l}\right>$ all contain the product of four 
$U$ matrix elements: 
each up-type fermion (quark or lepton) gives a factor
of $\chi_u^{\dagger}U\,u_L = (U_{21},U_{22})u_L$; each down-type fermion
produces a factor $\chi_d^{\dagger}U\,d_L = (-U_{11},-U_{12})d_L$.
The resulting integrals are straightforward, and most of them
vanish.  The only non-zero combinations are
\bea
U_{11}^2U_{22}^2 &\to& 2\pi^2/3,\nonumber\\
(U_{12}U_{21})^2 &\to& 2\pi^2/3,\\
U_{11}U_{22}U_{12}U_{21}&\to& -\pi^2/3.\nonumber
\eea
Because of this, the only non-zero Green's functions are 
\be
\bar{u}\bar{u}\bar{d}\bar{e},\phantom{AAA}\mbox{and}
\phantom{AAA}\bar{u}\bar{d}\bar{d}\bar{\nu},
\ee
and therefore conserve $U(1)_{em}$.
Adding $SU(3)_c$ indices, there are six independent
Green functions:
\be
\begin{array}{cc}
\bar{u}^1\bar{u}^2\bar{d}^3\bar{e},
&\:\:\:\:\bar{u}^1\bar{d}^2\bar{d}^3\bar{\nu},\\
\bar{u}^1\bar{d}^2\bar{u}^3\bar{e},
&\:\:\:\:\bar{d}^1\bar{u}^2\bar{d}^3\bar{\nu},\\
\bar{d}^1\bar{u}^2\bar{u}^3\bar{e},
&\:\:\:\:\bar{d}^1\bar{d}^2\bar{u}^3\bar{\nu}.\\
\end{array}
\ee
These all come in with the same sign because of the ordering
of the zero mode integrations in the functional integral.
They all have the same numerical prefactor, as well.

  Consider now the Green's function for 
$\bar{u}^1\bar{u}^2\bar{d}^3\bar{e}$.  The corresponding
amplitude is proportional to
\bea
&&\int dU\;(U_{21},U_{22})u_L^1\;(U_{21},U_{22})u_L^2\;
(U_{11},U_{12})d_L^3\;(U_{11},U_{12})e_L\\
&=&\frac{2\pi^2}{3}\left[u_{L_1}^1u_{L_1}^2d_{L_2}^3e_{L_2}
+u_{L_2}^1u_{L_2}^2d_{L_1}^3e_{L_1} - \frac{1}{2}
(u_{L_1}^1u_{L_2}^2 + u_{L_2}^1u_{L_1}^2)
(d_{L_1}^3e_{L_2} +  d_{L_2}^3e_{L_1})\right],\nonumber
\eea
where $u_L, d_L, e_L$ denote the external polarization vectors,
and the lower indices are spinorial.
This amplitude can be reproduced at lowest order 
by adding to the low-energy effective Lagrangian the operator
\bea
&& \left[u_{L_1}^1u_{L_1}^2d_{L_2}^3e_{L_2}
+u_{L_2}^1u_{L_2}^2d_{L_1}^3e_{L_1} - \frac{1}{2}
(u_{L_1}^1u_{L_2}^2 + u_{L_2}^1u_{L_1}^2)
(d_{L_1}^3e_{L_2} +  d_{L_2}^3e_{L_1})\right],\nnmb\\
&=& \frac{1}{2}(u_L^1\cdot e_L)(u_L^2\cdot d_L^3) -
\frac{1}{2}(u_L^1\cdot d_L^3)(u_L^2\cdot e_L),
\eea
where now the $u_L, d_L$ and $e_L$ represent the field operators,
and in the last line we have re-expressed the operator in a manifestly
Lorentz-invariant form.
  
  It should also be possible to connect up the color indices
with an $\epsilon^{abc}$ tensor since the effective operator is expected to
be invariant under $SU(3)_c$.  Notice that
\be
\epsilon^{abc}u^au^bd^c = 2(u^1u^2d^3 + u^1d^2u^3 + d^1u^2u^3).
\ee
Therefore, we can combine all the $uude$ operators into
\bea
&&\frac{1}{2}\epsilon^{abc}\frac{1}{2}\left[
(u_L^a\cdot e_L)(u_L^b\cdot d_L^c) - (u_L^a\cdot d_L^c)
(u^b_L\cdot e_L)\right]
\nonumber\\
&=& \frac{1}{2}\epsilon^{abc}\;(u_L^a\cdot e_L)(u^b_L\cdot d^c_L).
\eea
Exactly the same thing can be done for the $udd\nu$ operators.

  Putting everything together, the effective four-fermion operator 
corresponding to a single $SU(2)_1$ anti-instanton is
\bea
\mathcal{O}_{\rm{eff}} &=& \frac{C}{g_1^8}\,e^{-8\pi^2/g_1^2(\mu)}
\left(\frac{1}{4\pi^2}\right)^{b_0/2-1}2^{b_0/2}
\left(\frac{\mu}{{\vv}}\right)^{b_0}
\,\Gamma(1+b_0/2)\,\left(\frac{\pi^2}{3V_g}\right)\cdot
\nnmb\\
&&\phantom{AAAAAAAAA}\cdot\left(\frac{1}{{\vv}^2}\right)\,\epsilon^{abc}\,
\left[(u_L^a\cdot e_L)(u^b_L\cdot d^c_L) 
+ (d_L^a\cdot \nu_L)(d^b_L\cdot u_L^c)\right].\phantom{{A\choose A}}
\label{eqn:oeff}
\eea
where $V_g = 8\pi^2$ is four times the group volume,
$b_0$ is the one-loop beta-function coefficient, 
${\vv} \simeq \sqrt{2}\,u$, and the constant $C$ is given in
Eq.~(\ref{prefactor}).  This operator is also invariant
under $SU(2)_L$, and violates both $B$ and $L$ by one unit each.

%%%%%%%%%%%%%%%%%%%%%%%%%%%%%%%%%%%%%%%%%%%%%%%%%%%%%%%
%%%%%%%%%%%%%%%%%%%%%%%%%%%%%%%%%%%%%%%%%%%%%%%%%%%%%%%

\section{$B+L$-Violating Scattering by $SU(2)_1$ Instantons}
\label{scatt1}

  As a first application of the results of Section~\ref{inst},
we compute the scattering cross section for $bb\to \bar{t}\bar{\nu}$
due to $SU(2)_1$ instantons.  We will focus on this particular
process because of all the $B+L$ violating reactions induced
by the operator in Eq.~(\ref{eqn:oeff}), this one is expected to 
have the largest cross section at the LHC.  To see why,
note that this operator involves only third generation fermions.
As a result, when the parton-level cross section is convolved with 
parton distribution functions
(PDF's) to obtain the total hadronic cross section, it will be
suppressed by the small PDF's of the third-generation fermions
within the proton.  This suppression is fairly strong for the 
bottom quark, but extremely strong for the top quark.
Therefore, events with only bottom quarks in the initial state
are expected to produce the largest cross sections.

  The parton level cross section is computed straightforwardly
using the operator from Eq.~(\ref{eqn:oeff}).  Inserting
the $\epsilon^{abc}(b_L^a\cdot\nu_L)(b_L^b\cdot t^c_L)$
operator in the corresponding matrix element, and squaring, 
summing, and averaging over spins and colors, we find
\be
\epsilon^{abc}(b_L^a\ccdot\nu_L)(b_L^b\ccdot t^c_L) \to
\frac{2}{3}\left[2(p_1\ccdot p_3)(p_2\ccdot p_4)
+ 2(p_1\ccdot p_4)(p_2\ccdot p_3) - (p_1\ccdot p_2)(p_3\ccdot p_4)\right],
\ee
where $p_1$ and $p_2$ are the incoming momenta,
and $p_3$ and $p_4$ are the outgoing momenta.
The parton-level cross section then follows in the usual way.
To get the total cross section in a $pp$ hadron collider such 
as the LHC, we must convolve this cross section 
with the bottom quark PDF's of the proton.  Thus
\be
\sigma_{tot} = \int_0^1dx_1\int_0^1dx_2\;f_{{b}}(x_1)
f_{{b}}(x_2)\;\sigma(s=x_1x_2\,s_0),
\ee
where $\sqrt{s_0}$ is the $pp$ center-of-mass (CM) energy.
%We used CTEQ6M PDF's from Ref.~\cite{cteq6} in our analysis.
Since the bottom quark PDF's peak at small $x$, 
a large CM energy is needed to avoid a strong additional
suppression of the total cross section.

\begin{figure}[htb]
\centerline{
        \includegraphics[width=0.7\textwidth]{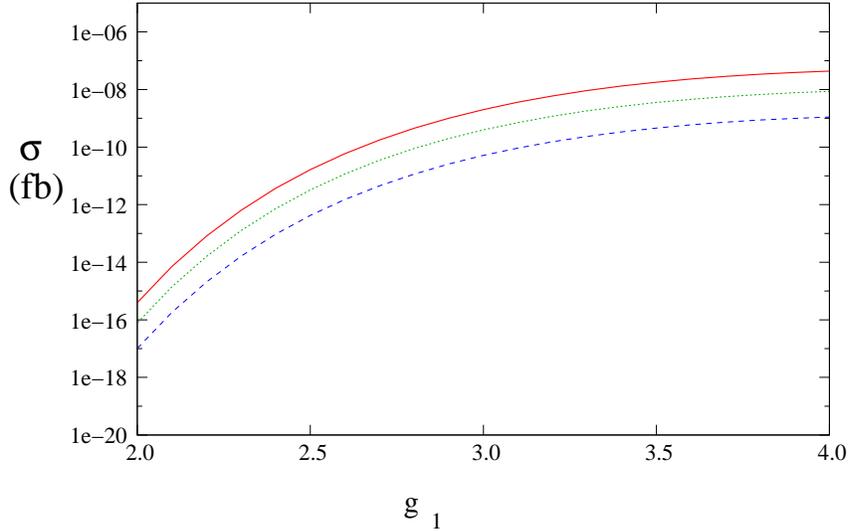}}
        \caption{The $SU(2)_1$ instanton mediated 
${b}{b}\to\bar{t}\bar{\nu}_{\tau}$ cross section at $\sqrt{s_0} = 14$~TeV 
for $u = 2$~TeV (solid red), $u=3$~TeV (dotted green), and $u = 5$~TeV (dashed blue).}
\label{cross}        
\end{figure}

  Fig.\ \ref{cross} shows the cross section for 
$bb\to\bar{t}\bar{\nu}_{\tau}$ scattering at the LHC, 
with $\sqrt{s_0} = 14$~TeV.  The three lines in this figure 
correspond to three different values of the $\su12$ symmetry
breaking VEV: $u = 2$~TeV, 3~TeV, and 5~TeV.
The CTEQ6M parton distributions from Ref.~\cite{cteq6} were
used to evaluate Eq.~(\ref{cross}).
Unfortunately, this $B+L$ violating cross section is 
unobservably small at the LHC, even for larger values
of the gauge coupling.  The reason why may be understood
by examining the various factors that contribute to the instanton
amplitude of Eq.~(\ref{eqn:coeffa}).  For $g_1 \simeq 3$, the 
usual instanton term, $e^{-8\pi^2/g_1^2}$, is still fairly small,
and there is an additional suppression by the $1/g_1^8$ term 
in the amplitude.  Together, they contribute a factor of
order $10^{-8}$.  This is offset somewhat by the large prefactor
$C$, given in Eq.~(\ref{prefactor}), which is of order $10^5$ in
the present case, but not enough for the cross section to be 
observable.  We would also like to emphasize that for very
large values of the gauge coupling $g_1$, the semi-classical
approximation used to derive the effective instanton operator
is expected to break down.

\section{Proton Decay from $SU(2)_1$ Instantons}
\label{pdecay1}

  The observed stability of the proton often leads to 
very strong constraints on theories beyond the Standard
Model which contain baryon number violating interactions.
This is true for the $\su12$ extension considered here
since the operator of Eq.~(\ref{eqn:oeff}) violates $B$ and 
$L$ by one unit, and can induce the decay of the proton
into a meson and a light lepton.  As we shall see below,
the experimental limit on the proton lifetime implies a 
\emph{lower} bound on the $SU(2)_1$-breaking 
scale $u$, and an \emph{upper} bound on the gauge coupling $g_1$.      

\begin{figure}[hbt!]
\begin{minipage}[t]{0.47\textwidth}
        \includegraphics[width = \textwidth]{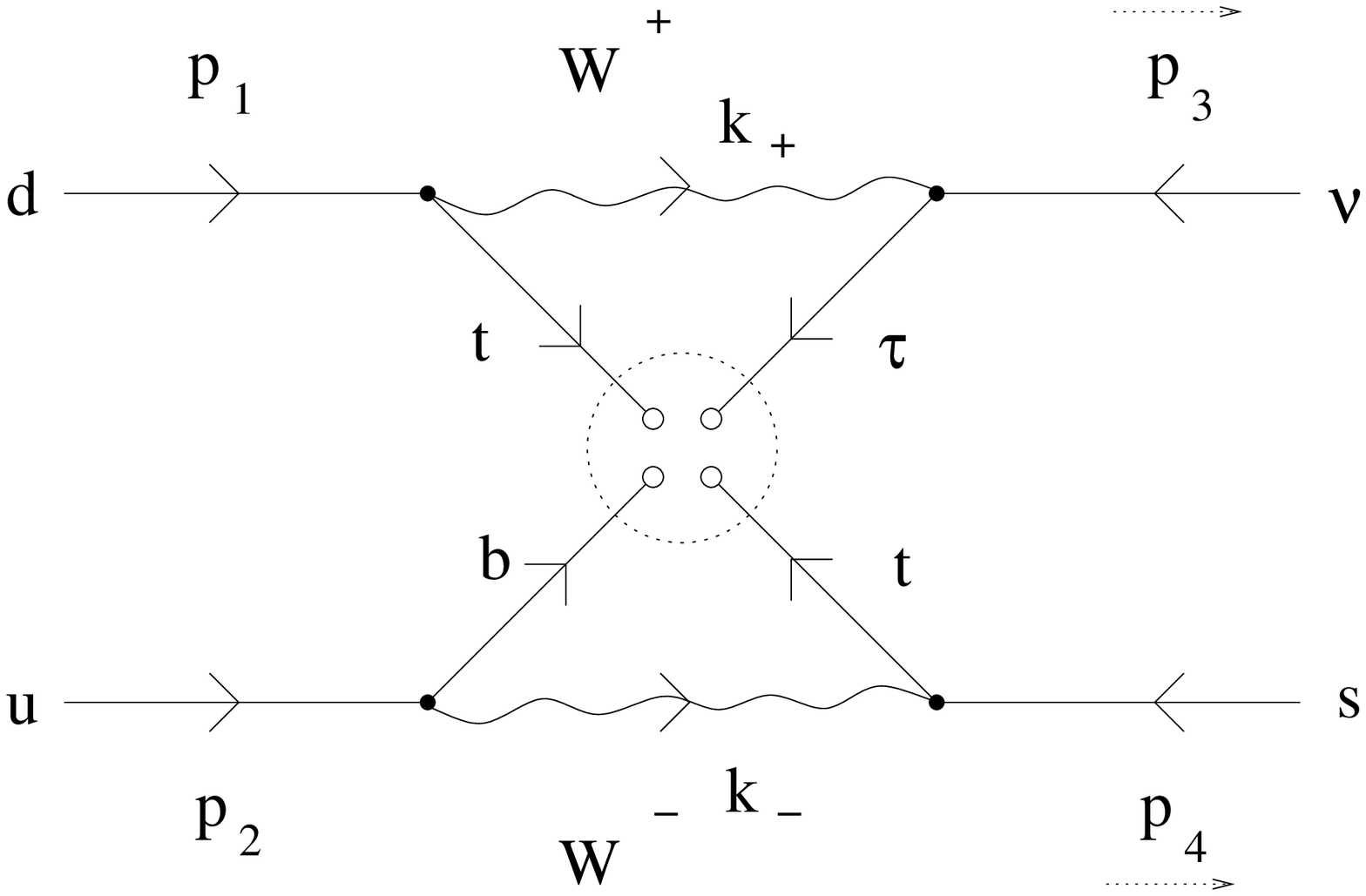}
\end{minipage}
\phantom{aa}
\begin{minipage}[t]{0.47\textwidth}
        \includegraphics[width = \textwidth]{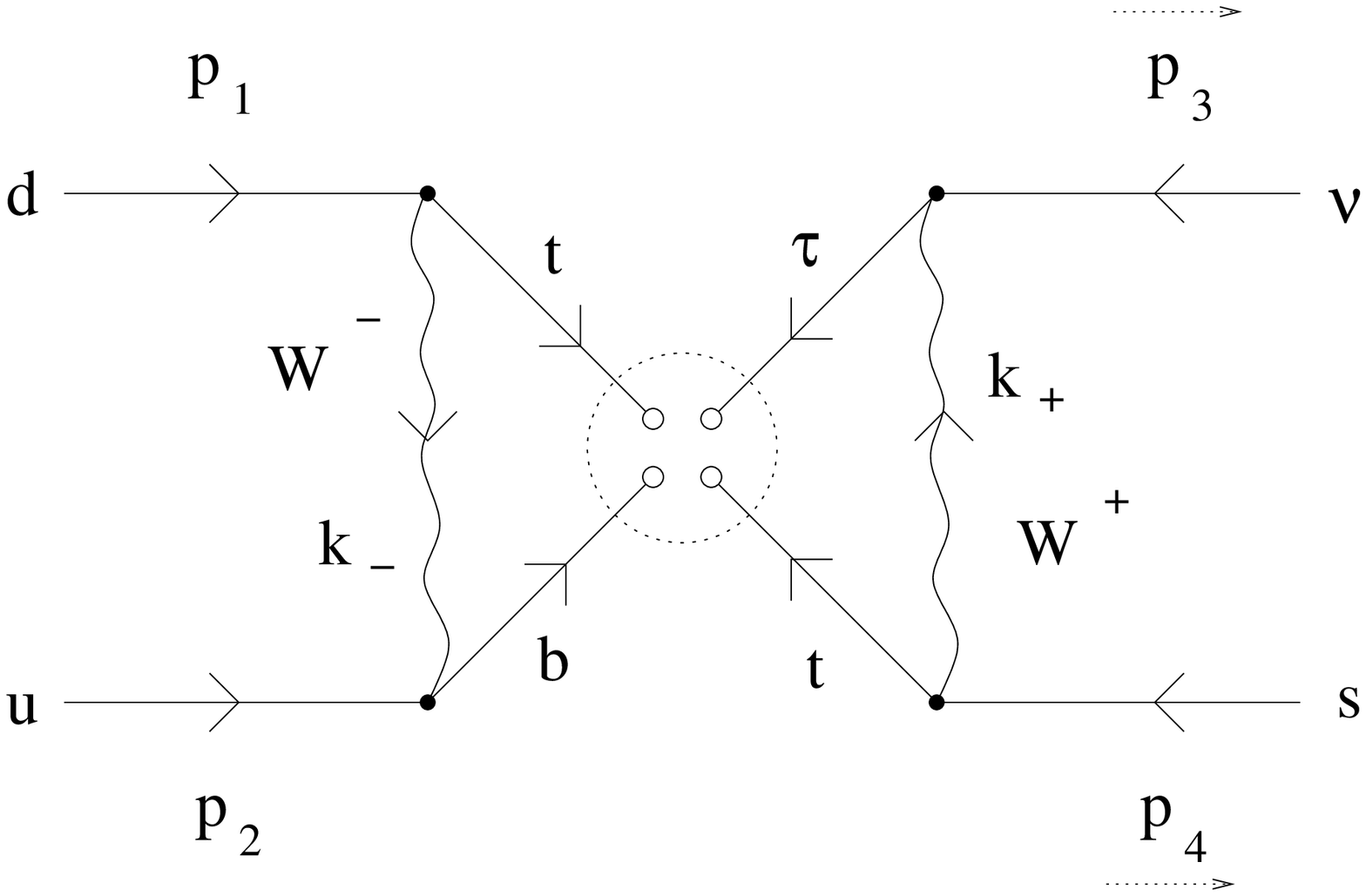}
\end{minipage}
\caption{Feynman diagrams for anti-instanton 
mediated proton decay.}
\label{diagrams}
\end{figure}

   For $SU(2)_1$ instanton induced decays to occur, however, 
the third generation quarks must be connected with the first 
generation quarks that make up the proton.  Such a link is 
provided by the flavor-changing couplings of the quarks with 
the W gauge bosons.  The Feynman diagrams for the process
$p\to K^+\bar{\nu}_{\tau}$ generated in this way are shown in 
Fig.\ \ref{diagrams}.
Both of these are suppressed by two loop factors.  
A second possibility, that avoids this loop suppression,
is that the light quark mass eigenstates in the proton
contain a small admixture of the third generation gauge eigenstates
that couple directly to $SU(2)_1$.  This generates a contribution
to the proton decay amplitude that is not suppressed by any 
loop factors, but does involve elements of the up and down 
quark mixing matrices.  Since these elements are unknown
(only their product is measured through the CKM matrix),
we will ignore this possibility and focus solely on the 
contributions involving $W$ boson loops.  Barring unusual 
cancellations, this will set a lower bound on the instanton-induced
proton decay rate.

  The operator responsible for $p\to K^+\bar{\nu}_{\tau}$ decay
is the $\epsilon^{abc}(t_L^a\ccdot b_L^b)(b_L^c\ccdot \tau_L)$
term in Eq.~(\ref{eqn:oeff}).  By connecting the legs of this
operator to first and second generation quarks through $W$ bosons,
as shown in Fig.\ \ref{diagrams}, we obtain a pair of operators
that directly mediate proton decay.  Both of these diagrams involve 
a pair of loop integrations, and in each case the two loops
are independent as a result of the locality of the effective operator.

  The loop integrals all have the form   
\be
I_{\mu\nu} = \int\frac{d^4k}{(2\pi)^4}\frac{(p_a-k)_{\mu}}
{[(p_a-k)^2-m_a^2]}\frac{(p_b-k)_{\nu}}{[(p_b-k)^2-m_b^2]}
\frac{1}{k^2-M_W^2},
\label{loopint}
\ee
where $p_a$ and $p_b$ are the external momenta, and $m_a$ and $m_b$
are the fermion masses in the loop.
This integral is logarithmically divergent in the ultraviolet.
The reason for this apparent divergence is that we have used 
the long-distance form of the fermion zero modes, 
which go as $p_{\mu}/p^2$, as shown in Eq.~(\ref{fzero}).  
For scales above $\rho^{-1}$, however, this form is no longer valid, 
and should
be replaced by the Fourier transform of the short-distance form
for the zero modes.  From Eq.~(\ref{eqn:zmshort}), we find that these go as
\be
x_{\mu}f(r) = \frac{x_{\mu}}{r(r^2+\rho^2)^{3/2}},
\ee
where $r=(|x^2|)^{1/2}$. The Fourier transform can be computed using
the identity
\be
\int d^4x e^{ip\cdot x}x_{\mu}f(r) = 4\pi^2\,i
\frac{p_{\mu}}{p^2}\int_0^{\infty}dr J_2(pr)r^3f(r).
\ee
For large $p$, $J_2(pr) \simeq \sqrt{\frac{2}{\pi\,pr}}\,\cos(pr-5\pi/4)$.
The resulting $r$ integral is finite, and the momentum-space 
wavefunction falls off at least as fast as $p^{-3/2}$ for large $p$.
Using this form in the loop integration at large momenta, 
the full integral is found to be convergent. Taking this fact into account,
we will approximate the result of the loop integrals, Eq.~(\ref{loopint}),
by cutting them off at a scale $\Lambda \sim \rho^{-1} \sim 
\sqrt{2}\pi\,u$, where our effective operator description is 
expected to break down.

  Setting the external momenta $p_a$ and $p_b$ to zero
in Eq.~(\ref{loopint}) and performing the integration, we find
\be
I_{\alpha\beta} = \eta_{\alpha\beta}\;
\frac{i}{8(2\pi)^2}\int_0^1\!dx\,dy\,dz\,\delta(1-x-y-z)\,
\left[\ln\left(1+\frac{\Lambda^2}{\Delta}\right)-\frac{3}{2}\right]
+ \mathcal{O}\left(\frac{\Delta}{\Lambda^2}\right),
\ee
with $\Delta$ given by
\be
\Delta = x\,m_a^2 + y\,m_b^2 + z\,M_W^2 + \mathcal{O}(p_1^2,p_2^2).
\ee
The integrals over $x$, $y$, and $z$ can be done analytically,
and the result is
\bea
I_{\alpha\beta} &=& \eta_{\alpha\beta}\,A(m_a^2,m_b^2,M_W^2)\\
\label{loop}
&=& \eta_{\alpha\beta}\;\frac{i}{16(2\pi)^2}
\left[-\frac{1}{2} + f_{\Lambda}(m_a^2,m_b^2,M_W^2)
+ f_{\Lambda}(M_W^2,m_a^2,m_b^2) + f_{\Lambda}(m_b^2,M_W^2,m_a^2)\right]
%+\mathcal{O}\left(\frac{\Delta}{\Lambda^2}\right)
,\nnmb
\eea
where
\be
f_{\Lambda}(a,b,c) = \frac{a^2}{(a-b)(a-c)}\left[
\ln\left(\frac{\Lambda^2}{a}\right)+\frac{1}{2}\right].
\ee

   The operators generated by the diagrams of 
Fig.\ \ref{diagrams} are found to be
\be
\mathcal{O}_{\rm{eff}} = -\left(\frac{24\pi^2}{3V_g}\right)\,V_f\,I_f\,L_f\,
\epsilon^{abc}\left[(u_L^a\ccdot s_L^b)(d_L^c\ccdot \nu_{L}^{\tau})    
+ (u_L^a\ccdot d_L^b)(s_L^c\ccdot \nu_{L}^{\tau})\right],
\label{pdop}
\ee
where $V_f$ is the product of W vertex factors, $L_f$ is the product
of the loop factors, and $I_f$ comes from the instanton prefactor.
The vertex factor is 
\be
V_f = \left(\frac{g}{\sqrt{2}}\right)^4\,V_{ts}V^*_{ub}V_{td}
\ee
The loop factor was computed above, and is given by
\be
L_f = A(m_t^2,m_b^2,M_W^2)\,A(m_t^2,m_{\tau}^2,M_W^2),
\ee
where the function $A$ is defined in Eq.~(\ref{loop}).
Finally, the instanton factor is the prefactor
of Eq.~(\ref{eqn:oeff}), and has the value
\be
I_f = \frac{C}{g_1^8}e^{-8\pi^2/g_1^2{\mu}}\left(\frac{\mu}
{\vv}\right)^{b_0}(4\pi^2)^{1-b_0/2}2^{b_0/2}
\Gamma(1+b_0/2)\;\frac{1}{\vv^2}\,
\label{if}
\ee
with the constant $C$ given by Eq.~(\ref{prefactor}).

  The matrix elements of the operators in
Eq.~(\ref{pdop}) between $p$ and $K^+$ states are given 
in~\cite{Aoki:1999tw}.  They are
\bea
\epsilon^{abc}\bra{K^+}(u_L^as_L^b)d_L^c
\ket{p\phantom{K^+\hspace{-0.5cm}}} &=& 
\frac{\beta}{f_{\pi}}\frac{2m_p}{3 m_B}D\,P_L\,u_p\\
\epsilon^{abc}\bra{K^+}(u_L^ad_L^b)s_L^c
\ket{p\phantom{K^+\hspace{-0.5cm}}} &=& 
\frac{\beta}{f_{\pi}}\left[1+(F+\frac{1}{3}D)
\frac{m_p}{m_B}\right]P_Lu_p.\nnmb 
\eea
Here, $u_p$ is a Dirac spinor for the external proton,
$f_{\pi}=0.131$~GeV is the pion decay constant,
$m_p = 0.94$~GeV is the proton mass, and $m_B = 1.15$~GeV
is an average baryon mass.  The parameters $F\simeq 0.44$
and $D\simeq 0.81$ come from converting the quark operator
to baryons and mesons via chiral perturbation theory. The parameter
$\beta = 0.014(1)$~GeV$^3$ is computed on the lattice
in~\cite{Aoki:1999tw}.\footnote{Based on previous estimates
of this quantity, however, there may be up to an order of magnitude
systematic uncertainty in its value.}

  The Dirac spinor for the proton gets contracted 
(using $\epsilon^{\alpha\beta}$) with the Dirac spinor for
the neutrino.  
%Doing this, and summing/averaging over spins
%we get 
%\be
%\frac{1}{2}\sum_{s,\,s'}\left|u_{p_L}(p_0)_{\alpha}\,
%\epsilon^{\alpha\beta}\,\nu_L(p_3)_{\beta}\right|^2 = p_0\cdot p_3.
%\ee
After summing and averaging over spins, we find the decay rate
%Inserting this into the formula for the decay rate, we find
\be
\Gamma(p\to K^+\bar{\nu}_{\tau}) =
\frac{(m_p^2-m_K^2)^2}{32\pi\,m_p^3f_{\pi}^2}|\mathscr{A}|^2
\ee
where $\mathscr{A}$ is given by
\be
\mathscr{A} = \beta\left(\left[1+(F+\frac{1}{3}D)
\frac{m_p}{m_B}\right] + \frac{2}{3}D\frac{m_p}{m_B}\right) 
\cdot V_f\,L_f\,I_f\,\left(\frac{24\pi^2}{3V_g}\right),
\label{aterm}
\ee
where $V_f$, $L_f$, and $I_f$ are given above.

In computing the numerical value of the proton decay rate, 
we set the renormalizaton scale in Eq.~(\ref{if}) equal to the 
symmetry breaking scale, $\mu = \vv$.  This corresponds to a matching
at this scale.  In principle, one should also include the
running of the effective operator induced by QCD.  However,
we ignore this effect, as it is expected to be of order unity.

\begin{figure}[hbt]
\begin{center}
  \includegraphics[width = 0.7\textwidth]{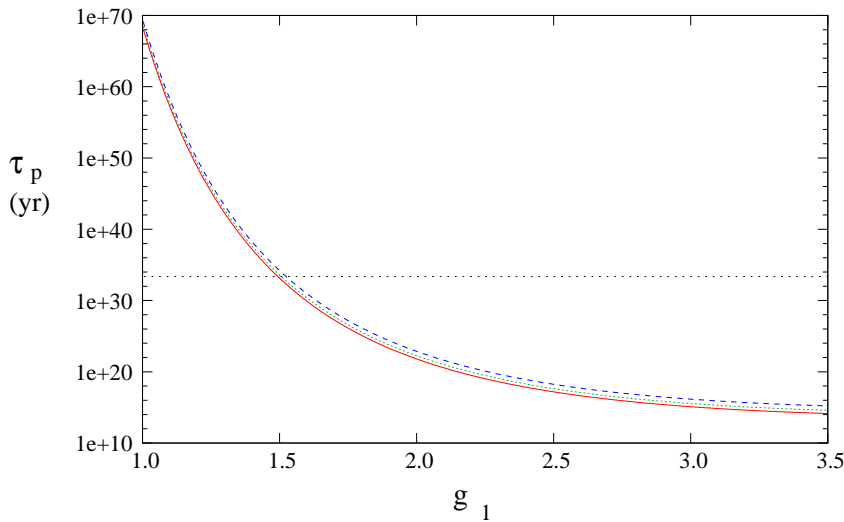}
\end{center}
  \caption{Proton lifetime due to $SU(2)_1$ instantons 
for $u= 2$~TeV (solid red), $u = 3$~TeV (dotted green), and $u = 5$~TeV (dashed blue).
Also shown in this figure (flat dotted line) is the $90\%$~c.l. 
experimental lower bound on the proton lifetime~\cite{Kobayashi:2005pe}.}
\label{taup}
\end{figure}

The instanton mediated proton lifetime as a function of the 
$SU(2)_1$ coupling is shown in Fig.~\ref{taup}.
%\footnote{The conversion
%factor is $\tau_p = \left(\frac{\mbox{GeV}}{\Gamma}\right)
%\,2.087\times 10^{-32}yr$.}
Also shown in this figure is the current experimental $90\%$~c.l. 
limit on proton decay via $p\to K^+\bar{\nu}$~\cite{Kobayashi:2005pe}:
\be
\tau_p > 2.3\times 10^{33}\,yr.
\ee
From the figure, we see that $g_1\lesssim 1.5$ is required
to satisfy the proton decay constraint.  This upper limit
on the gauge coupling $g_1$ puts an interesting bound
on models that make use of the $\su12$ gauge structure,
such as topflavor and non-commuting extended technicolor.
It also limits the amount by which the Higgs mass may be raised
through $D$-terms in supersymmetric theories.  

The results above were obtained for values of $u$ of the order of a
few TeV. The bounds on $g_1$ may be relaxed by increasing the value
of $u$. However, since the proton decay rate is proportional to 
$u^{-4}$, while it depends exponentially on the value of $g_1^{-2}$,
a large increase on $u$ would be necessary to significantly modify 
the bounds on $g_1$. Alternatively, one can find a lower bound on
$u$ for a particular value of $g_1$. For instance, for a value
of $g_1 \simeq 2.5$, the bound on $u$ is found to be $u \simgt 10^{8}$~GeV. 
The large value of the lower bound on $u$ reflects the relatively 
mild dependence on this parameter.  We have also assumed that the 
effective quartic coupling for the symmetry breaking bidoublet
field is small, $\lambda \ll 2\pi^2$.  For larger values of $\lambda$,
as sometimes arise in technicolor-type models~\cite{Hill:2002ap}, 
there will be an additional suppression of the instanton amplitude 
leading to a longer proton lifetime for given values of $g_1$ and $u$.

As we will see below, the bounds from nucleon decay
are significantly weakened if there are additional fermions,
beyond the third generation of the SM, that are charged under 
$SU(2)_1$.  These could arise, for instance, as the superpartners
of the Higgs scalars in a supersymmetric theory 
or from additional exotic quarks or leptons.

\section{Strongly-Coupled Light Fermions}
\label{light}

  In the previous sections we have discussed the effects of
instantons of the $SU(2)_1$ gauge group when its coupling 
becomes large.  Since this group couples only to the third
generation, one of these effects is the generation of
four-fermion operators.  
One such operator, that of Eq.~(\ref{pdop}), leads to 
the rapid decay of the proton if the gauge coupling $g_1$ is 
too large.  This implies an upper bound on $g_1$ (for a given $u$)
that provides a relevant constraint on several models making
use of the $\su12$ gauge structure.  This operator also generates
$B+L$ violating scattering events in particle colliders, but
unfortunately the cross section for these is too small to be
observed at the LHC, especially given the upper bound on $g_1$.
%unfortunately precludes the possibility of having
%observable $SU(2)_1$ instanton mediated $B+L$-violating events 
%at the LHC.  
A second possibility, the one we consider 
in this section, is that the gauge coupling of the $SU(2)_2$ group 
%which couples to the fermions of the first and second generations, 
becomes large.  In this case, it is the $SU(2)_2$ instantons 
that become unsuppressed, possibly leading to observable effects.

Since the first and second generations of fermions couple to $SU(2)_2$,
the effective operators generated by the $SU(2)_2$ instantons
will involve \emph{eight} fermions, violate $B$ and $L$ by two
units each, and will be accompanied by a factor of $u^{-8}$.  
These operators can therefore mediate dinucleon decay,
the limits on which are nearly as stringent as those for 
proton decay.  However, because of the $u^{-8}$ factor,
the decay rates will be suppressed by a factor of 
$(m_p/u)^{16}$, which is of order $10^{-50}$ for $u\sim \mbox{TeV}$.
On the other hand, the scattering cross sections mediated by
the instanton will go as $(\sqrt{s}/u)^{16}$.  As up or a down quarks 
with energies of order 1~TeV can be found with non-vanishing probability
at the LHC, this prefactor is not exceedingly small.
Indeed, the PDF's for valence quarks at high energies
are much larger than for the bottom, which provides an additional
enhancement compared to the previous case.

\subsection{Di-Nucleon Decay}

  Using the results of Eq.~(\ref{eqn:coeffa}) and Section~\ref{inst},
the eight-fermion operators generated by $SU(2)_2$ instantons
will have the form
\bea
\mathcal{O}_{\rm{eff}} &=& \frac{C}{g_2^8}\,e^{-8\pi^2/g_2^2}
\left(\frac{1}{4\pi^2}\right)^{b_0/2}2^{3+b_0/2}\Gamma(4+b_0/2)
\left(\frac{\mu}{\vv}\right)^{b_0}\frac{1}{\vv^8}\,\tilde{\mathcal{O}},
\nnmb\\
&:=& \phantom{AA} \frac{\tilde{C}}{\vv^8}\,\tilde{\mathcal{O}}
\label{olight}
\eea
where $C$ is given in Eq.~(\ref{prefactor}), and 
$\tilde{\mathcal{O}}$ is a linear combination of 
$(uude)(ccs\mu)$, $(uude)(ssc\nu_{\mu})$,
$(ddu\nu_e)(ccs\mu)$, and $(ddu\nu_e)(ssc\nu_{\mu})$.  
These operators all have $B=L=2$, and can therefore
induce the decay of a pair of nucleons.  

\begin{figure}[htb]
\centerline{
        \includegraphics[width=0.55\textwidth]{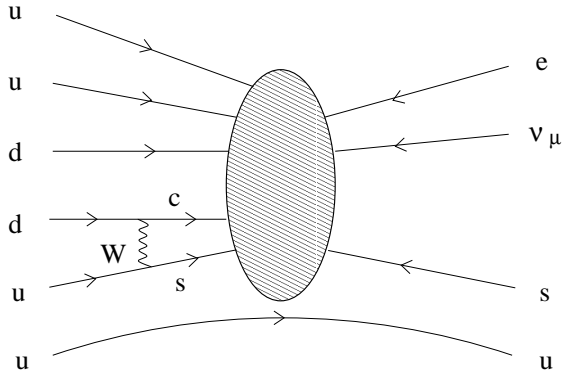}}
        \caption{A Feynman diagram for diproton decay
due to an $SU(2)_2$ instanton.}
\label{pdecayl}        
\end{figure}

  We will consider the di-proton decay rate induced by the operator 
$(uude)(ssc\nu_{\mu})$.  The relevant Feynman diagram with the least
possible number of loops is shown in Fig.\ \ref{pdecayl}.
Calculating the amplitude for this diagram is complicated because
of the nuclear physics uncertainties associated with the overlap
of the proton wave functions.  To make an estimate of the amplitude,
we shall simply replace all unknown dimensionful terms by the proton
mass $m_p$.  This is likely a gross overestimate of the decay rate,
especially since the relevant nuclear physics scale is closer
to $1~\rm{fm}^{-1} \sim 0.2$~GeV, so our results should be considered as
a robust upper bound on the actual rate.  
With this approximation, the di-proton lifetime is given by
\be
\Gamma \simeq |\tilde{C}|^2\left(\frac{g}{\sqrt{2}}\right)^4\,
\left|A(m_c^2, m_s^2, M_W^2)\right|^2\,|V_{us}V^*_{cd}|^2
\,\left(\frac{m_p}{\vv}\right)^{16}\frac{1}{m_p},
\label{ratepp}
\ee
where the function $A(a,b,c)$ was defined in Eq.~(\ref{loop}).
As for the proton decay rate, we match the effective operator
at scale $\mu = \vv$, and neglect the running below this scale.
 
\begin{figure}[htb]
\centerline{
        \includegraphics[width=0.7\textwidth]{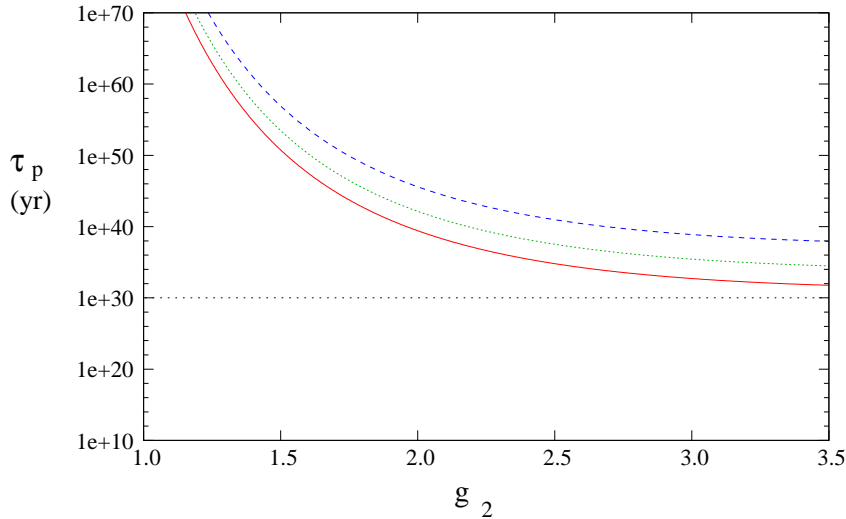}}
        \caption{The di-proton lifetime induced by 
$SU(2)_2$ instantons.  In this figure, the solid blue line
corresponds to $u= 2$~TeV, the dashed green line is for $u=3$~TeV,
and the dotted blue line is for $u = 5$~TeV.
The flat dotted line is the current experimental lower bound.}
\label{pdlight}        
\end{figure}

  The current best experimental limit on di-nucleon decay processes
was obtained by the Fr\'ejus collaboration, which looked for 
di-nucleon decay in iron, and is of the order of 10$^{30}$ years. 
The corresponding di-proton lifetime, obtained from our estimate
of Eq.~(\ref{ratepp}), is shown in Fig.\ \ref{pdlight}. 
The estimated lifetime is many orders of magnitude above the experimental
bound, even for very large values of the $SU(2)_2$ coupling.
As noted above, the additional suppression relative to the $SU(2)_1$
case comes from the factor of $(m_p/\vv)^{16}$ in Eq.~(\ref{ratepp}).
Thus, the experimental limit on the $pp$ lifetime does not impose 
any strong constraint on the coupling constant $g_2$.

  Another possible effect of the $B=L=2$ operators considered in
this section are hydrogen--antihydrogen oscillations, as first
suggested by Feinberg, Goldhaber and Steigman~\cite{Feinberg:1978sd}.
Observe that, neglecting CP-violation, the existence
of $\Delta B = \Delta L = 2$ interactions determines that the
real mass eigenstates of Hydrogen are
\begin{equation}
H_1 = \frac{1}{\sqrt{2}} \left( H \pm \bar{H} \right)
\end{equation}
and will have a small mass difference. Oscillations between
a pure hydrogen and antihydrogen states will occur 
with a period $T = 2\pi/\Delta m$, that, due to astrophysical
bounds must be larger than $7\times 10^{10}$ years. 
However, the dominant, instanton mediated process violate 
baryon and lepton number but also flavor. Therefore, these transitions
are suppressed not only by the small instanton amplitude
and $(m_p/\vv)^{16}$, but also by loop and mixing angle factors. 
A simple examination of the relevant factors involved in the 
baryon number violating transition suggests that the mass difference 
induced by the baryon number violating is much larger than 
the experimental bound ($T >   10^{40}$ years), and is therefore 
unmeasurably small.  Finally, we also note that neutron oscillations 
are not induced by the instanton operators because they also violate 
lepton number by two units.

\subsection{Scattering by $SU(2)_2$ Instantons}

  Contrary to the case in which only one generation couples to
the strongly interacting sector, the baryon number violating
processes occurring in proton-proton collisions at the
LHC involve six quarks and two leptons. In the following,
we shall consider the scattering of two first generation quarks 
leading to a final state with four energetic jets and two 
first and second generation \emph{same-sign} leptons. 
This is a spectacular event with very little background 
in the standard model, and can be easily detected when 
the two outgoing leptons are charged. 

  As in the previous subsection, the large number of fermion
legs makes a precise calculation very difficult, so we will only
estimate the relevant scattering cross section.  
In particular, we will focus on the operator $(uude)(ccs\mu)$,
which can induce $uu \to \bar{d} {e}^+ \bar{c}\bar{c}\bar{s} \mu^+$
at the parton level.  This particular channel is the most promising
one for two reasons.  First, the $uu$ initial state is the most
probable with respect to the PDF's of the proton, and second,
the two charged like-sign leptons in the final state produce 
a distinctive signature for these events.  
We also note that this cross section is enhanced by the fact that
the LHC is a $pp$ collider, and not a $p\bar{p}$ collider such
as the Tevatron, since the instanton-mediated scattering events
involve two particles instead of a particle and an anti-particle.  

  The scattering amplitude induced by the $(uude)(ccs\mu)$ operator
has the form
\be
\mathcal{A} = \frac{\tilde{C}}{\vv^8}\,\bar{h},
\ee
where $\tilde{C}$ is the factor defined in Eq.~(\ref{olight})
and $\bar{h}$ is the matrix element of the $(uude)(ccs\mu)$ term
between the external states.  The cross section derived from 
this amplitude is 
\be
\sigma = \frac{1}{s}|\tilde{C}|^2\,\left[\prod_{i=3}^8\int 
\frac{d^3k_i}{(2\pi)^32E_i}\right]\,(2\pi^4)\delta^{(4)}(p_1+p_2
-p_3-\ldots -p_8)\,|\bar{h}|^2,
\ee
in which $|\bar{h}|^2$ includes summation and averaging over spin
and color states.  To proceed, we must approximate the phase
space integral.  For this, we shall assume that 
\be
|\bar{h}|^2 \sim \left(\frac{\sqrt{s}}{2}\right)^2\prod_{i=3}^8E_i,
\ee
since in the leading term, each fermion is expected to 
contribute a factor of its momentum.
Using the methods of~\cite{kbyck}, we find that
\bea
&&\left[\prod_{i=1}^n\int 
\frac{d^3k_i}{(2\pi)^32E_i}\right]\,(2\pi^4)\,\delta^{(4)}(p_1+p_2
-p_3-\ldots -p_8)\,\prod_{i=1}^nE_i\nnmb\\ 
&=& \frac{1}{2}(4\pi)^{3-2n}\frac{1}{(\frac{3}{2}n-1)!(\frac{3}{2}n-2)!}
s^{3n/2-2},
\eea
valid for large $n$.  Our estimate for the (parton-level) 
cross section is therefore
\be
\sigma = \frac{1}{s}\,|\tilde{C}|^2\frac{1}{8}(4\pi)^{-9}\frac{1}{7!\,8!}
\left(\frac{\sqrt{s}}{\vv}\right)^{16}.
\label{sigmal} 
\ee

\begin{figure}[htb]
\centerline{
        \includegraphics[width=0.7\textwidth]{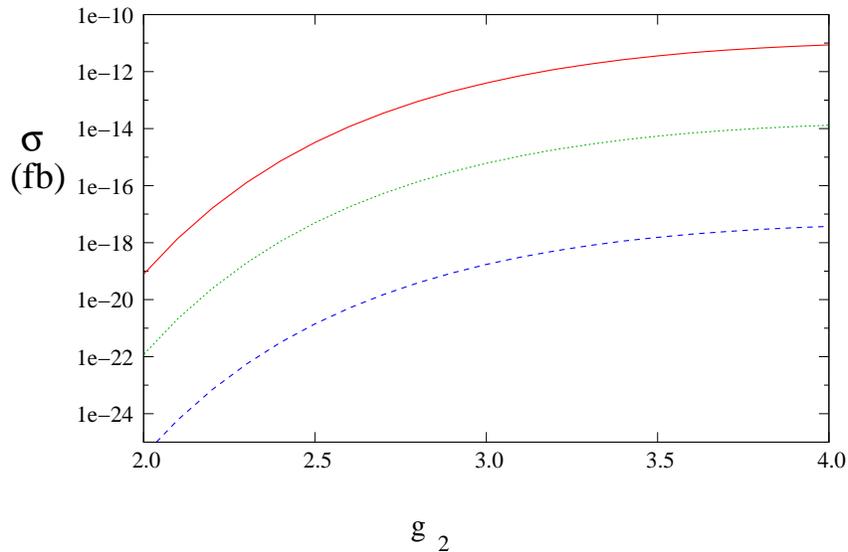}}
        \caption{The instanton mediated 
cross section at $\sqrt{s_0} = 14$~TeV for the case
in which the first two generations are charged under
the strong $SU(2)_1$ interactions, for  three values of the scale 
$u = 2$~TeV (solid red), 3~TeV (dotted green), and 5~TeV (dashed blue).}
\label{cross2}        
\end{figure}

  As in Section~\ref{scatt1}, this cross section must be convolved with
the $u$ quark PDF's in order to get the full cross section.
Doing so, we find the total cross sections shown in Fig.~\ref{cross2}
for a center-of-mass energy of $\sqrt{s_0} = 14$~TeV.
Like the cross sections due to $SU(2)_1$ instantons, these cross sections
are too small to be observed at the LHC.  Different from the $SU(2)_1$
case, however, the $SU(2)_2$ cross section is not suppressed by 
a small instanton prefactor ($\tilde{C}$ defined in
Eq.~(\ref{olight}) is of order unity for $g_2 \sim 3$) or the product
of bottom quark PDF's.  Instead, the very small phase space factor of 
Eq.~(\ref{sigmal}) is responsible for inhibiting the instanton events.
These results are also very sensitive to the value of 
$u \simeq \vv/\sqrt{2}$ and the center of mass energy $\sqrt{s}$
due to the high power of $\vv$ and $\sqrt{s}$ appearing
in the cross section expression. 

\begin{comment}
The characteristic energy of an up- or down-quark in a proton-proton
collision at the LHC is of the order of 1 to 2 TeV, and therefore
the $u$ suppression factor is much weaker than in the case of 
proton decay. However, there is a strong phase space suppression
that combined with the mild $u$ suppression and the exponential
form factor suppression still leads to small cross sections for
values of $u$ of the order of a few TeV. Therefore, even 
for values of $\alpha$ of order one, these 
processes, although spectacular, will be rare at the LHC. The
appearance of equal sign electrons and a muons in the final state,
and the absence of missing energy in the underlying event make
these processes background free, beyond instrumental or
misidentification effects. 
\end{comment}

~\\
~\\
%%%%%%%%%%%%%%%%%%%%%%%%%%%%%%%%%%%%%%%%%%%

\section{Conclusions}
\label{conc}

In this article we have shown that the rates of anomalous
$B+L$ violating transitions in gauge extended models can be
much larger than in the SM.  For models based on the group
$\su12$, such as topflavor and non-commuting extended technicolor,
we have found that the instanton mediated scattering 
cross sections are too small to be observed at the LHC,
but that nucleon decay implies an upper bound on the $SU(2)_1$
gauge coupling.  This limit is relevant for these models, and may (through
dimensional deconstruction) provide a glimpse into some non-perturbative
processes relevant for certain five dimensional theories.
It similarly suggests that the possibility of raising the Higgs mass 
through $D$-terms in supersymmetric theories is limited by the bound
on the $SU(2)_1$ gauge coupling.
The opposite limit has the $SU(2)_2$ felt by the first and second generations
to be strongly interacting.
However, the instantonic effects associated with the 
$SU(2)_2$ gauge group are generally too weak to be seen, 
even for large values of the gauge coupling. The rate of baryon and
lepton number violating processes are strongly suppressed by
the small phase space factors arising in this  case.

As a byproduct of this analysis, 
we have also re-examined the constraints on the $\su12$ gauge
structure implied by the precision electroweak data.  Our results are
roughly in agreement with those in the literature.
In general, we find that to agree with the data, the symmetry breaking
scale of the extended gauge group must be greater than a few TeV, although
the limits may be relaxed in the case that only the third generation 
fermions are coupled to the strongly interacting gauge group.

It may be possible that other types of experiments could be sensitive to
extended gauge instantons.  For example, even higher energy colliders such
as a VLHC will see less suppression and could have observable rates if the
integrated luminosity is sufficiently high.  Also, it is possible that 
horizontal air showers induced by cosmic neutrinos could be detected
by cosmic ray observatories.  Furthermore, they may open a new avenue for
electroweak-style baryogenesis.  While these topics are all beyond the scope
of the present work, they are interesting possibilities and show that
non-perturbative effects from new interactions may be just as exciting and
important as the perturbative effects.

%%%%%%%%%%%%%%%%%%%%%%%%%%%%%%%%%%%%%%%%%%%%%%%%%%%%%%%%%%%%%%%%%
~\\
~\\
~\\
{\bf Acknowledgments}

Work at ANL is supported in part by the US DOE, Div.\ of HEP, Contract 
W-31-109-ENG-38. C.\ Wagner would like to thank N.\ Weiner and S.\ Chivukula
for useful discussions and comments.  T.\ Tait has benefitted 
from discussions with B.\ Dobrescu and C.\ Hill.  D.\ Morrissey 
would like to thank C.\ Bal\'azs, P.\ Batra, and C.\ Hill 
for several helpful conversations.

\newpage

\begin{appendix}

\section{Euclidean Space Spinor Conventions\label{apspin}}

  We use the following conventions in Minkowski space:
\bea
\eta^{\mu\nu} &=& diag(+1,-1,-1,-1),\\
\sigma^{\mu} &=& (\sigma^{\mu})_{\alpha\dot{\alpha}} 
= (\mathbb{I},\vec{\sigma}),\nonumber\\
\bar{\sigma}^{\mu} &=& (\bar{\sigma}^{\mu})^{\dot{\alpha}{\alpha}} 
= (\mathbb{I},-\vec{\sigma}),\nonumber\\
\sigma^{\mu\nu} &=& \frac{i}{4}(\sigma^{\mu}\bar{\sigma}^{\nu}
-\sigma^{\nu}\bar{\sigma}^{\mu}),\nonumber\\
\bar{\sigma}^{\mu\nu} &=& \frac{i}{4}(\bar{\sigma}^{\mu}{\sigma}^{\nu}
-\bar{\sigma}^{\nu}{\sigma}^{\mu}).\nonumber
\eea

  In Euclidean space we take our vectors to be 
\bea
p_4 &=& -ip_0,\nonumber\\
p^e_{\mu} &=& (p_i,p_4).
\eea
and define the Euclidean space $\sigma$-matrices according to
\bea
\sigma^e_{\mu} &=& (\vec{\sigma},i),\\
\bar{\sigma}^e_{\mu} &=& (\vec{\sigma},-i),\nonumber\\
\sigma^e_{\mu\nu} &=& \frac{1}{4i}(\sigma^e_{\mu}\bar{\sigma}^e_{\nu}
-\sigma^e_{\nu}\bar{\sigma}^e_{\mu}),\nonumber\\
\bar{\sigma}^e_{\mu\nu} &=& \frac{1}{4i}(\bar{\sigma}^e_{\mu}
{\sigma}^e_{\nu}-\bar{\sigma}^e_{\nu}{\sigma}^e_{\mu}).\nonumber
\eea
This is slightly different from the conventions in Ref.~\cite{espinosa}.

With these definitions, it follows that
\bea
v_{\mu}w^{\mu} &=& -v^e_{\mu}w^e_{\mu},\nonumber\\
v_{\mu}\sigma^{\mu} &=& v^e_{\mu}\sigma^e_{\mu},\nonumber\\
v_{\mu}\bar{\sigma}^{\mu} &=& -v^e_{\mu}\bar{\sigma}^e_{\mu}\nonumber\\
v_{\mu}w_{\nu}\sigma^{\mu\nu} &=& v^e_{\mu}w^e_{\nu}\sigma^e_{\mu\nu},\\
v_{\mu}w_{\nu}\bar{\sigma^{\mu\nu}} &=& v^e_{\mu}w^e_{\nu}
\bar{\sigma}^e_{\mu\nu}\nonumber,
\eea
where repeated lower indices are summed over.

In terms of the 't Hooft symbols, $\eta_{a\mu\nu}$,~\cite{thooft}
we have
\bea
\sigma_{\mu\nu}^e &=& \bar{\eta}_{a\mu\nu}\sigma^a/2,\\
\bar{\sigma}_{\mu\nu}^e &=& {\eta}_{a\mu\nu}\sigma^a/2,\nonumber
\eea
where $a = 1,2,3$ is an $SU(2)$ index.  The $e$'s will
be left implicit in most of the expressions in this work.   
We will also follow the convention of Ref.~\cite{espinosa}
and use $\sigma$'s for spin-space sigma-matrices, and $\tau$'s for
the $SU(2)$-space sigma-matrices.

%%%%%%%%%%%%%%%%%%%%%%%%%%%%%%%%%%%%%%%%%%%%%%%%%%%%%%%
%%%%%%%%%%%%%%%%%%%%%%%%%%%%%%%%%%%%%%%%%%%%%%%%%%%%%%%

\section{Gauge Bosons in the $\su12$ Model}
\label{apew}
%%%%%%%%%%%%%%%%%%%%%%%%%%%%%%%%%%%%%%%%%%%%

We list here the gauge boson masses and couplings
in the $\su12$ light and heavy gauge extensions.  In both cases, the
gauge coupling for the light set of weak bosons is related to the two
original $SU(2)$ gauge couplings by,
\be
g_L = \frac{g_1g_2}{\sqrt{g_1^2+g_2^2}} ~.
\ee
To simplify expressions, we introduce the short-hand notation,
\bea
c_{\varphi} \equiv \cos\varphi &=& \frac{g_1}{\sqrt{g_1^2+g_2^2}}~,\\
s_{\varphi} \equiv \sin\varphi &=& \frac{g_2}{\sqrt{g_1^2+g_2^2}}~, \nnmb
\eea
for the $SU(2) \times SU(2)$ gauge couplings, and
\bea
s_{\theta} \equiv \sin\theta &=& \frac{g_y}{\sqrt{g_y^2+g_L^2}} ~, \\
c_{\theta} \equiv \cos\theta &=& \frac{g_L}{\sqrt{g_y^2+g_L^2}} ~, \nonumber
\eea
is the analog of the weak mixing angle in the SM.

\subsection{The \emph{Heavy} Case}

The charged gauge boson states consist of 
$A_j^{\pm} = (A_j^1\mp i A_j^2)/\sqrt{2}$, $j=1,2$.
In this basis, the mass matrix reads
\be
M_{\pm}^2 =
u^2\,\left(\begin{array}{cc}
g_1^2(1+\delta)&-g_1g_2\\
-g_1g_2&g_2^2
\end{array}\right),
\ee
where $\delta \equiv v^2/2u^2$.  By assumption, $\delta \ll 1$, and
we treat it as a perturbation, keeping only the terms necessary to compute 
the leading order in $\delta$ to EW observables.

The mass eigenstates, $W$ and $W^\prime$, 
are related to these, to $\mathcal{O}(\delta)$, by the transformation
\be
\left(\begin{array}{c}A_1\\A_2\end{array}\right)=
\left(\begin{array}{cc}
\sf - \sf\cf^4\delta & -\cf-\sf^2\cf^3\delta\\
\cf + \sf^2\cf^3\delta & \sf - \sf\cf^4\delta
\end{array}\right)
\left(\begin{array}{c}W\\W^\prime\end{array}\right) ~.
\ee
and the charged gauge boson masses are given by
\bea
M^2_{W} &=& \frac{g_L^2\,v^2}{2}\left(1 - c^4_{\varphi}\delta\right)
\; , \\
M^2_{W^\prime} &=& \frac{g_L^2\,u^2}{s_{\varphi}^2c_{\varphi}^2}
\:=\: (g_1^2+g_2^2)\,u^2\; \nonumber
{\label{mw}}
\eea
where, as above, $g_L = g_1\,g_2/\sqrt{g_1^2+g_2^2}$ is the gauge
coupling of the diagonal $SU(2)_L$ subgroup.

The coupling of these gauge bosons to the fermions of the first
and second generations has the form 
\be
g_2\,A_2 \to g_L(1+\sf^2\,\cf^2\,\delta)\,W 
+ g_L(\frac{\sf}{\cf}-\sf\cf^3\,\delta)\,W^\prime,
\label{g12w}
\ee
while the coupling with the third generation fermions is given by
\be
g_1\,A_1 \to g_L(1 - \cf^4\,\delta)\,W 
+ g_L(-\frac{\cf}{\sf}-\sf\cf^3\,\delta)\,W^\prime.
\label{g3w}
\ee

The mass matrix for the neutral states in basis $(B,A_1,A_2)$ is given by
\be
M_0^2 = u^2\,\left(\begin{array}{ccc}
g_y^2\,\delta&-g_1g_y\,\delta&0\\
-g_1g_y\,\delta&g_1^2 \left(1 + \,\delta \right)&-g_1g_2\\
0&-g_1g_2&g_2^2
\end{array}\right) ~.
\ee
The transformation to the mass eigenstates, $(A,Z,Z^\prime)$ has the form
\be
\left(\begin{array}{c}B\\A_1\\A_2\end{array}\right) =
\left(\begin{array}{ccc}
\ct&-\st&\frac{\st}{\ct}\sf\cf^3\,\delta\\
\sf\st&\sf\ct - \frac{\sf\cf^4}{\ct}\,\delta&-\cf-\sf^2\cf^3\,\delta\\
\cf\st&\cf\ct+\frac{\sf^2\cf^3}{\ct}\,\delta&\sf-\sf\cf^4\,\delta
\end{array}\right)
\left(\begin{array}{c}A\\Z\\Z^\prime\end{array}\right) ~.
\ee
The masses of the $Z$ and $Z^\prime$ are 
\bea
{\label{mz}}
M_Z^2 &=& \frac{g_L^2\,v^2}{2\,\ct^2}(1-\cf^4\,\delta)\\
M_{Z^\prime}^2&=& (g_1^2+g_2^2)\,u^2.\nonumber
\eea
The couplings of the first and second generations are 
\be
(g_2\,A_2t^3 + g_y\,Y\,B) \:\to\:
e\,Q\,A + \frac{g_L}{\ct}\left[(t^3-Q\,\st^2) 
+ \sf^2\cf^2\delta\,t^3\right]\,Z
+ g_L\frac{\sf}{\cf}\,t^3\,Z^\prime ,
\label{g12z}
\ee
where $Q = (t^3+Y)$, as usual, and for the third generation we have
\be
(g_1\,A_1\,t^3 + g_y\,Y\,B) \:\to\:
e\,Q\,A + \frac{g_L}{\ct}\left[(t^3-Q\,\st^2) 
- \cf^4\delta\,t^3\right]\,Z
- g_L\frac{\cf}{\sf}\,t^3\,Z^\prime .
\label{g3z}
\ee

\subsection{The \emph{Light} Case}

The analysis of the light case is very similar to the 
previous section.
The charged gauge boson mass matrix, in basis $(A_1, A_2)$, is
\be
M_{\pm}^2 =
u^2\,\left(\begin{array}{cc}
g_1^2&-g_1g_2\\
-g_1g_2&g_2^2(1+\delta)
\end{array}\right),
\ee
where, again, $\delta = v^2/2u^2 \ll 1$.  
The corresponding mixing matrix is,
\be
\left(\begin{array}{c}A_1\\A_2\end{array}\right)=
\left(\begin{array}{cc}
\sf + \sf^3\cf^2\delta & -\cf+\sf^4\cf\delta\\
\cf - \sf^4\cf\delta & \sf + \sf^3\cf^2\delta
\end{array}\right)
\left(\begin{array}{c}W\\W^\prime\end{array}\right) ~ ,
\ee
and the charged gauge boson masses are given by
\bea
M^2_{W} &=& \frac{g_L^2\,v^2}{2}\left(1 - s^4_{\varphi}\delta\right),\\
M^2_{W^\prime} &=& \frac{g_L^2\,u^2}{s_{\varphi}^2c_{\varphi}^2}
\:=\: (g_1^2+g_2^2)\,u^2\; .\nonumber
{\label{mwl}}
\eea
The coupling of these gauge bosons to the fermions of the first
and second generations has the form 
\be
g_2\,A_2 \to g_L(1-\sf^4\,\delta)\,W 
+ g_L(\frac{\sf}{\cf}+\sf^3\cf\,\delta)\,W^\prime,
\label{g12wl}
\ee
while the coupling with the third generation fermions is given by
\be
g_1\,A_1 \to g_L(1 + \sf^2\cf^2\,\delta)\,W 
+ g_L(-\frac{\cf}{\sf}+\sf^3\cf\,\delta)\,W^\prime.
\label{g3wl}
\ee
  
The mass matrix for the neutral states, in the basis
$(B,A_1,A_2)$, is given by
\be
M_0^2 = u^2\,\left(\begin{array}{ccc}
g_y^2\,\delta&0&-g_2g_y\,\delta\\
0&g_1^2&-g_1g_2\\
-g_2g_y\,\delta&-g_1g_2&g_2^2 \left( 1+\delta \right)
\end{array}\right).
\ee
leading to the 
transformation to the mass eigenstates $(A,Z,Z^\prime)$,
\be
\left(\begin{array}{c}B\\A_1\\A_2\end{array}\right) =
\left(\begin{array}{ccc}
\ct&-\st&-\frac{\st}{\ct}\sf^3\cf\,\delta\\
\sf\st&\sf\ct + \frac{\sf^3\cf^2}{\ct}\,\delta&-\cf+\sf^4\cf\,\delta\\
\cf\st&\cf\ct-\frac{\sf^4\cf}{\ct}\,\delta&\sf+\sf^3\cf^2\,\delta
\end{array}\right)
\left(\begin{array}{c}A\\Z\\Z^\prime\end{array}\right) ~,
\ee
with $Z$ and $Z^\prime$ masses, 
\bea
{\label{mzl}}
M_Z^2 &=& \frac{g_L^2\,v^2}{2\,\ct^2}(1-\sf^4\,\delta)\\
M_{Z^\prime}^2&=& (g_1^2+g_2^2)\,u^2.\nonumber
\eea
The first and second generation couplings are 
\be
(g_2\,A_2t^3 + g_y\,Y\,B) \:\to\:
e\,Q\,A + \frac{g_L}{\ct}\left[(t^3-Q\,\st^2) 
- \sf^4\delta\,t^3\right]\,Z
+ g_L\frac{\sf}{\cf}\,t^3\,Z^\prime + \mathcal{O}(\delta^2),
\label{g12zl}
\ee
and the third generation couplings are,
\be
(g_1\,A_1\,t^3 + g_y\,Y\,B) \:\to\:
e\,Q\,A + \frac{g_L}{\ct}\left[(t^3-Q\,\st^2) 
+ \sf^2\cf^2\delta\,t^3\right]\,Z
- g_L\frac{\cf}{\sf}\,t^3\,Z^\prime + \mathcal{O}(\delta^2).
\label{g3zl}
\ee

\section{Precision Electroweak Constraints}
\label{app2}

Using the results of the previous appendix, 
we perform the matching to input parameters and compute the
shifts in the electroweak observables in both the 
\emph{heavy} and \emph{light} gauge-extended models.  In both cases,
$\alpha$ has the same form as in the SM:
\be
\alpha = \frac{e^2}{4\,\pi} = \frac{g_L^2\,\sin^2\theta}{4\,\pi} ~,
\ee
and $g_L^2$ is given by,
\bea
g_L^2 &=& \frac{4\,\pi\,\alpha}{\sin^2\theta} ~.
\eea

\subsection{\emph{Heavy} Case}

The expression for $M_Z$ is given in Eq.~(\ref{mz}):
\be
M_Z^2 = \frac{g_L^2\,v^2}{2 c^2_{\theta}}(1-\cf^4\,\delta).
\ee
For $G_F$, which is extracted from muon decay, we must consider
the low-energy effective four-fermion couplings which arise 
from integrating out both the $W$ and $W^\prime$ bosons.  
Using the charged gauge boson
masses, Eq.~(\ref{mw}), as well as their couplings to the 
first and second generation fermions, we find
\bea
4\sqrt{2}\,G_F &=& \left[\frac{g_L^2\,v^2}{2}\left(1 - c^4_{\varphi}\,
\delta\right)\right]^{-1}\;g_L^2(1+\sf^2\cf^2\,\delta)^2
+ \left[\frac{g_L^2\,u^2}{s_{\varphi}^2c_{\varphi}^2}\right]^{-1}
\;g_L^2\left(\frac{\sf}{\cf}\right)^2\nonumber\\
&=&\frac{2}{v^2}(1 + \delta).
\eea

Inverting these relations, we match to our input parameters,
\bea
v^2 &=& \frac{(1+\delta)}{2\sqrt{2}\,G_F}\\
\sin^2\theta &=& \frac{1}{2}-\frac{1}{2}
\sqrt{1-4\,A_0[1+(1-\cf^4)\delta]},\nonumber
%\sqrt{1-4\,A_0})
%+ \frac{A_0}{\sqrt{1-4\,A_0}}(1-\cf^4\,)\delta,\nonumber
\eea
where
\be
A_0 = \frac{\pi\,\alpha}{\sqrt{2}\,G_F\,M_Z^2} \simeq 0.179059 ~.
\ee

These are sufficient to work out the shifts in many of the 
electroweak observables relative to the SM.  The important ones for our
analysis are,
\bea
M_W&=&(M_W)_{SM}\left[1-0.219(1-\cf^4)\delta \right]\nonumber\\
&&\nonumber\\
\Gamma_Z&=&(\Gamma_Z)_{SM}\left[1+\left(
-1.348+0.790\cf^4+1.684\sf^2\cf^2\right)\,\delta\right]\nonumber\\
\Gamma_{had}&=&(\Gamma_{had})_{SM}\left[1+\left(
-1.478+0.974\cf^4+1.828\sf^2\cf^2\right)\,\delta\right]\nonumber\\
\Gamma_{e,\mu}&=&(\Gamma_{e,\mu})_{SM}\left[1+\left(
-1.175+1.175\cf^4+2.122\sf^2\cf^2\right)\,\delta\right]\nonumber\\
\Gamma_{inv}&=&(\Gamma_{inv})_{SM}\left[1+\left(
-1.000+0.333\cf^4+1.333\sf^2\cf^2\right)\,\delta\right]\nonumber\\
&&\nonumber\\
R_{b}&=&(R_{b})_{SM}\left[1+\left(
\:\:0.059-1.846\cf^4-1.828\sf^2\cf^2\right)\,\delta\right]\nonumber\\
R_{c}&=&(R_{c})_{SM}\left[1+\left(
-0.114+0.618\cf^4+0.583\sf^2\cf^2\right)\,\delta\right]\nonumber\\
R_{\tau}&=&(R_{\tau})_{SM}\left[1+\left(
-0.302+1.921\cf^4+1.828\sf^2\cf^2\right)\,\delta\right]\nonumber\\
R_{e,\mu}&=&(R_{e,\mu})_{SM}\left[1+\left(
-0.302-0.201\cf^4-0.293\sf^2\cf^2\right)\,\delta\right]\nonumber\\
&&\nonumber\\
A_{b}&=&(A_{b})_{SM}\left[1+\left(
-0.232+0.071\cf^4\right)\,\delta\right]
\nonumber\\
A_{c}&=&(A_{c})_{SM}\left[1+\left(
-1.786+1.786\cf^4+1.242\sf^2\cf^2\right)\,\delta\right]\nonumber\\
A_{s}&=&(A_{s})_{SM}\left[1+\left(
-0.232+0.232\cf^4+0.161\sf^2\cf^2\right)\,\delta\right]\nonumber\\
A_{\tau}&=&(A_{\tau})_{SM}\left[1+\left(
-20.391+6.215\cf^4\right)\,\delta\right]
\nonumber\\
A_{e,\mu}&=&(A_{e,\mu})_{SM}\left[1+\left(
-20.391 +20.391\cf^4+14.17\sf^2\cf^2\right)\,\delta\right]\nonumber\\
&&\nonumber\\
A^{b}_{FB}&=&(A^b_{FB})_{SM}\left[1+\left(
-20.621+20.462\cf^4+14.17\sf^2\cf^2\right)\,\delta\right]\nonumber\\
A^{c}_{FB}&=&(A^c_{FB})_{SM}\left[1+\left(
-22.171+22.171\cf^4+15.41\sf^2\cf^2\right)\,\delta\right]\nonumber\\
A^{s}_{FB}&=&(A^s_{FB})_{SM}\left[1+\left(
-20.621+20.621\cf^4+14.333\sf^2\cf^2\right)\,\delta\right]\nonumber\\
A^{\tau}_{FB}&=&(A^{\tau}_{FB})_{SM}\left[1+\left(
-40.771+26.602\cf^4+14.17\sf^2\cf^2\right)\,\delta\right]\nonumber\\
A^{e,\mu}_{FB}&=&(A^{e,\mu}_{FB})_{SM}\left[1+\left(
-40.771+40.771\cf^4+28.34\sf^2\cf^2\right)\,\delta\right] %\nonumber\\
\eea

\subsection{The \emph{Light} Case}

The corresponding expressions for the \emph{light} case are
\bea
M_Z^2 &=& \frac{g_L^2\,v^2}{2\ct^2}(1-\sf^4\,\delta),\nnmb\\
4\sqrt{2}\,G_F &=& \frac{2}{v^2}.
\eea
These translate into
\bea
v^2 &=& \frac{1}{2\sqrt{2}\,G_F}\\
\sin^2\theta &=& \frac{1}{2}-\frac{1}{2}
\sqrt{1-4\,A_0(1-\sf^4\delta)}.\nonumber
\eea
The corresponding shifts in the electroweak observables
are
\bea
M_W&=&(M_W)_{SM}\left[1 + 0.219\sf^4 \delta \right]\nonumber\\
&&\nonumber\\
\Gamma_Z&=&(\Gamma_Z)_{SM}\left[1+\left(
-1.348+1.684\sf^2\cf^2-0.383\sf^4\right)\,\delta\right]\nonumber\\
\Gamma_{had}&=&(\Gamma_{had})_{SM}\left[1+\left(
0.504\sf^2\cf^2-0.351\sf^4\right)\,\delta\right]\nonumber\\
\Gamma_{e,\mu}&=&(\Gamma_{e,\mu})_{SM}\left[1+\left(
 -0.947\sf^4\right)\,\delta\right]\nonumber\\
\Gamma_{inv}&=&(\Gamma_{inv})_{SM}\left[1+\left(
0.667\sf^2\cf^2-0.333\sf^4\right)\,\delta\right]\nonumber\\
R_{b}&=&(R_{b})_{SM}\left[1+\left(
1.787\sf^2\cf^2+1.770\sf^4\right)\,\delta\right]\nonumber\\
R_{c}&=&(R_{c})_{SM}\left[1+\left(
-0.504\sf^2\cf^2 -0.469\sf^4\right)\,\delta\right]\nonumber\\
R_{\tau}&=&(R_{\tau})_{SM}\left[1+\left(
-1.618\sf^2\cf^2 -1.526\sf^4\right)\,\delta\right]\nonumber\\
R_{e,\mu}&=&(R_{e,\mu})_{SM}\left[1+\left(
0.504\sf^2\cf^2 +0.596\sf^4\right)\,\delta\right]\nonumber\\
&&\nonumber\\
A_{b}&=&(A_{b})_{SM}\left[1+\left(
0.161\sf^2\cf^2 +0.232\sf^4\right)\,\delta\right]\nonumber\\
A_{c}&=&(A_{c})_{SM}\left[1+\left(
0.545\sf^4\right)\,\delta\right]\nonumber\\
A_{s}&=&(A_{s})_{SM}\left[1+\left(
0.171\sf^4\right)\,\delta\right]\nonumber\\
A_{\tau}&=&(A_{\tau})_{SM}\left[1+\left(
14.171\sf^2\cf^2+ 20.386\sf^4\right)\,\delta\right]\nonumber\\
A_{e,\mu}&=&(A_{e,\mu})_{SM}\left[1+\left(
6.215\sf^4\right)\,\delta\right]\nonumber\\
&&\nonumber\\
A^{b}_{FB}&=&(A^b_{FB})_{SM}\left[1+\left(
0.161\sf^2\cf^2 +6.450\sf^4\right)\,\delta\right]\nonumber\\
A^{c}_{FB}&=&(A^c_{FB})_{SM}\left[1+\left(
6.760\sf^4\right)\,\delta\right]\nonumber\\
A^{s}_{FB}&=&(A^s_{FB})_{SM}\left[1+\left(
6.286\sf^4\right)\,\delta\right]\nonumber\\
A^{\tau}_{FB}&=&(A^{\tau}_{FB})_{SM}\left[1+\left(
14.171\sf^2\cf^2 +26.602\sf^4\right)\,\delta\right]\nonumber\\
A^{e,\mu}_{FB}&=&(A^{e,\mu}_{FB})_{SM}\left[1+\left(
12.431\sf^4\right)\,\delta\right]\nonumber\\
\eea

%\newpage

\end{appendix}

%%%%%%%%%%%%%%%%%%%%%%%%%%%%%%%%%%%%%%%%%%%%%%%%%%%%%%%%%%%%%%%%%

%\newpage

\end{document}